\documentclass[useAMS,usenatbib,fleqn]{mn2e}
\usepackage{times}
\usepackage{amsmath}
\usepackage{graphicx} 
\usepackage{multicol}
\usepackage{lastpage}

\usepackage{amssymb}
\usepackage{upgreek}
\usepackage{mathtools}
\usepackage{bm}

\def\D{{\rm D}}
\def\del{\upartial}

\def\grad_s{\nabla_{\! s}\,}
\def\grad_p{\nabla_{\! p}\,}
\def\bfx{\mathbfit{x}}

\def\bfv{\mathbfit{v}}
\def\bfk{\mathbfit{k}}
\def\frakp{\mathfrak{p}}

\def\gsim{\;\rlap{\lower 2.5pt\hbox{$\sim$}}\raise 1.5pt\hbox{$>$}\;}
\def\lsim{\;\rlap{\lower 2.5pt\hbox{$\sim$}}\raise 1.5pt\hbox{$<$}\;}

\def\gsim{\;\rlap{\lower 2.5pt\hbox{$\sim$}}\raise 1.5pt\hbox{$>$}\;}
\def\lsim{\;\rlap{\lower 2.5pt\hbox{$\sim$}}\raise 1.5pt\hbox{$<$}\;}
\def\del{{\upartial}}
\def\grad{\nabla}

\def\beq{\begin{equation}}
\def\eeq{\end{equation}}

\newcommand{\Del}{\bf \nabla}

\usepackage{color}
\definecolor{comred}{rgb}{.8,.2,0.1}
\definecolor{insgreen}{rgb}{.1,.5,0.1}

\title{Numerical Convergence of Hot-Jupiter Atmospheric Flow Solutions}

\author [J.~W.~Skinner \& J.~Y-K.~Cho] {J.~W.~Skinner,$^1$\thanks{Email:
    j.w.skinner@qmul.ac.uk} J.~Y-K.~Cho,$^2$\thanks{Email:
    jcho@flatironinstitute.org} \\ $^1$ School of Physics and Astronomy, Queen
  Mary University of London, Mile End Road, London E1 4NS, UK\\ $^2$ CCA,
  Flatiron Institute, 162 Fifth Ave, New York, NY, 10010, USA\\}

\begin{document}

\date{Accepted yyyy mmm dd. Received yyyy mmm dd; in original form yyyy mmm dd}

\pagerange{\pageref{firstpage}--\pageref{}} \pubyear{yyyy}

\maketitle

\label{firstpage}

\begin{abstract}

  We perform an extensive study of numerical convergence for hot-Jupiter
  atmospheric flow solutions in simulations employing a setup commonly-used in
  extrasolar planet studies -- a resting state thermally forced to a prescribed
  temperature distribution on a short time-scale at high altitudes.  Convergence
  is assessed rigorously with: {\it i})~a highly-accurate pseudospectral model
  which has been explicitly verified to perform well under hot-Jupiter flow
  conditions and {\it ii})~comparisons of the kinetic energy spectra,
  instantaneous (unaveraged) vorticity fields and temporal evolutions of the
  vorticity field from simulations which are numerically equatable.  In the
  simulations, the (horizontal as well as vertical) resolution, dissipation
  operator order and viscosity coefficient are varied with identical physical
  and initial setups.  All of the simulations are compared against a fiducial,
  reference simulation at high horizontal resolution and dissipation order (T682
  and $\Del^{\rm 16}$, respectively) -- as well as against each other.  Broadly,
  the reference solution features a dynamic, zonally (east--west) {\it
    a}symmetric jet with a copious amount of small-scale vortices and gravity
  waves.  Here we show that simulations converge to the reference simulation
  only at T341 resolution {\it and} with $\Del^{\rm 16}$ dissipation order.
  Below this resolution and order, simulations either do not converge or
  converge to unphysical solutions.  The general convergence behaviour is
  independent of the vertical range of the atmosphere modelled, from $\sim\!
  2\!\times\!  10^{-3}$\,MPa to $\sim\!  2\!\times\!  10^1$\,MPa.  Ramifications
  for current extrasolar planet atmosphere modelling and observations are
  discussed.

\end{abstract}


\section{Introduction}\label{intro}

A fundamental goal of a numerical method, as well as of a code implementing it,
is to generate a solution that approximates the true solution of the solved
equation(s) {\it and} whose approximation improves as the grid spacing or the
reciprocal of the truncation wavenumber tends to zero.  Accordingly, in
computational studies there is a long history of carefully assessing,
theoretically as well as empirically, the accuracy and convergence of numerical
schemes and codes
\citep[e.g.][]{GottOrsz77,Canuetal88,Boyd00,Stri04,Durr10,Lauretal11}.  In high
Reynolds number flows, such as those routinely encountered in astrophysics and
atmospheric physics, the flow field does not generally remain completely smooth
in time -- even when initialised with a smooth field.\footnote{A`smooth' field
  is a function that has many continuous derivatives.  More precisely, it is a
  function that meets the Lipschitz condition \citep[][]{Krey78}.}  Instead,
features develop with spatial scales close to, or at, the size of the individual
grid cell.  Such features cannot be accurately captured by any numerical method,
and the errors induced often feed back on the large scales -- thus exerting a
significant, deleterious influence on the overall solution.  In such
circumstances, there is not much recourse: one generally employs the finest grid
or the highest wavenumber truncation possible, along with a well-controlled
dissipation that eliminates the spurious poorly-resolved features while leaving
all of the well-resolved features unimpaired.  In light of this, the possibility
of a code erroneously converging to a `solution' that does not approximate the
true solution is a perennial concern in numerical studies \cite[see e.g.][and
  references therein]{Boyd00}.  This is especially so in fast-developing
research areas, such as extrasolar planets, wherein the flows modelled often
reside in a poorly-understood region of the dynamical parameter space and pose
severe computational challenges \citep[][]{Choetal15,Choetal19}.

At present, extrasolar planet atmospheric dynamics and general circulation
modelling studies commonly employ an idealised setup to generate flow and
temperature distributions starting from an initial state of rest
\citep[e.g.][]{Showetal08,Showetal09,Hengetal11,Bendetal13,LiuShow13,DobbAgol13,
  Maynetal14,Polietal14,Choetal15,Mendetal16,TanKoma19}.  Known as the
`Newtonian cooling approximation' \citep[see e.g.][]{Salb96,Choetal08}, the
setup consists of linearly `dragging' the flow temperature to a specified
temperature distribution on a specified time-scale at different pressure levels.
Although highly idealised, it is a reasonable and practical first representation
of the forcing in the absence of detailed information.  So far, a number of
studies have explored separately the effects of initial condition, numerical
resolution and explicit (as well as implicit) dissipation in hot-Jupiter
simulations using codes solving different equations with various resolutions,
algorithms and setups
\citep[e.g.][]{Choetal08,DobbLin08,Showetal09,ThraCho10,Hengetal11,ThraCho11,
  PoliCho12,Bendetal13,LiuShow13,DobbAgol13,Polietal14,Maynetal14,Choetal15,
  Mendetal18,Meno20} However, the issue of convergence under a controlled
setting with a rigorously-tested and numerically-accurate code {\it at high
  resolution} is still lacking.  That is, a clear and robust interpretation of
the simulation results, along with the true solution of the most basic setup, is
yet to be realised.

In this paper, we directly address this issue.  Here we report on the results
from a large number (over 300) of carefully prepared pseudospectral simulations
using the BOB code \citep{Rivietal02,Scotetal04,PoliCho12}, with equatable
simulations compared in several different ways.  While the discussion is focused
on hot-Jupiters and primitive equations,
we emphasize that the findings here are also relevant to other
tidally-synchronized objects (e.g. cool stars and telluric planets) as well as
the full Navier--Stokes equations; the latter have also been used in extrasolar
planet studies \citep[e.g.][]{DobbLin08,Maynetal14,Mendetal16}.  Note that the
primary difference between the primitive equations and the Navier--Stokes
equations is the assumption of hydrostatic balance in the former equations.
Formally, the hydrostatic assumption limits the validity of the primitive
equations to flow structures with a small aspect (vertical to horizontal) ratio.
The assumption also suggests a corresponding requirement for the ratio of
vertical to horizontal resolutions in numerical simulations; that is, if the
pressure scale height~${\cal H}$ is used as the vertical length scale, it
effectively sets ${\cal L} \ga 10{\cal H}$, where ${\cal L}$ is the horizontal
length scale.  For hot-Jupiters, this means roughly ${\cal L} \ga R_p/50$, where
$R_p$ is the planetary radius.\footnote{N.B. locally ${\cal H}$ can vary by a
  factor of $\sim\! 4$ in the modelled atmosphere.}  Hence, a spatial scale
range of roughly three orders of magnitude is covered by the primitive
equations.

\section{Methodology}\label{sec:method}

We solve the traditional primitive equations \citep[see e.g.][]{Salb96}.  In
$\bfx = (\lambda,\phi,p)$ coordinates representing (longitude, latitude,
pressure), the equations read:
\begin{subequations}\label{eq:pe}
  \begin{eqnarray}
    \frac{\D \bfv}{\D t}\ & = & -\grad_p \Phi - \Big(\frac{u}{R_p}\tan\phi+ f
    \Big)\bfk\times\bfv + {\cal D}_{\bfv} \\ \frac{\del\Phi}{\del p}\ & = &
    -\frac{1}{\rho} \\ \frac{\del\omega}{\del p}\ & = & -\grad_p\cdot\bfv
    \\ \frac{\D T}{\D t} & = & \frac{\omega}{\rho\, c_p} +
    \frac{\dot{q}_{\rm{net}}}{c_p} + {\cal D}_T\, ,
  \end{eqnarray}
\end{subequations}
where $\D / \D t \equiv \del / \del t + \bfv\!\cdot\!\grad_p + \omega\del / \del
p$ is the material derivative; $t$ is the time; $\bfv = (u,v)$ is the (eastward,
northward) velocity (in a frame rotating with rotation rate $\varOmega$) on a
constant $p$-surface; $\omega\equiv\D p / \D t$ is the `vertical' pressure
velocity in the rotating frame; $R_p$, the planetary radius, is fiducially set
to be at $p\! =\! 0.1$\,MPa $ (= 1$\,bar); $\bfk$ is the unit vector in the
local vertical direction; $\grad_p$ is the horizontal gradient on a constant
$p$-surface; $\Phi(\bfx,t) = gz(\bfx,t)$ is the geopotential, where $g$ is the
constant `surface gravity' at $z = R_p$ with $z$ the vertical distance above
$R_p$; $f(\phi) = 2 \varOmega\sin\phi$ is the Coriolis parameter, the projection
of the planetary vorticity vector~$2\bm{\mathit{\Omega}}$ onto~$\bfk$; the
direction of $\bm{\mathit{\Omega}}$ orients `north'; $T(\bfx,t)$ is the
temperature; $\cal{D}\!_{\chi}$, for $\chi \in \{\bfv, T\}$\footnote{In this
  paper, `$\{\,\cdot\,,\,\cdot\,,\,\ldots\}$', `$[\,\cdot\,,\cdot\,]$' and
  `$(\,\cdot\,,\,\cdot\,,\,\ldots)$' carry their usual meanings -- i.e., set,
  (closed) interval and tuple, respectively.}, are dissipations given by
\begin{eqnarray}\label{eq:hyper}
  {\cal D}\!_{\chi}\ =\ \nu_{2 \frakp}\big[(-1)^{\frakp+1}\grad\!_p^{\,2\frakp}
    + {\cal C}\big]\,\chi\, ,
\end{eqnarray}
where $\nu_{2\frakp}$ is the constant dissipation coefficient; $\frakp \in
\mathbb{Z}^+$ is the order of the dissipation (not to be confused with the
pressure $p$); ${\cal C} = (2/R_p^2)^{\frakp}$ is a term that compensates the
damping of uniform rotation by $\cal{D}\!_{\bfv}$, thus preserving angular
momentum conservation \citep[see e.g.][]{Polietal14}; $\rho(\bfx,t)$ is the
density; $c_p$ is the constant specific heat at constant pressure; and,
$\dot{q}_{\rm{\tiny net}}(\bfx,t)$ is the net diabatic heating rate.
Note that the $\frakp\! >\! 1$ instantiations of ${\cal D}\!_{\chi}$ are
known as `hyper-dissipation' \citep[e.g.][]{ChoPol96a,ThraCho11,PoliCho12}.
Broadly, hyper-dissipation has the effect of extending the inertial range by
focusing the energy dissipation rate to a narrow range of wavenumbers near the
truncation scale; this can be readily seen by taking the scalar product of
equation~(\ref{eq:pe}a) with $\bfv$ and then spectral transforming the
resulting equation.

Equations~(\ref{eq:pe}) are closed by the equation of state for an ideal gas, $p
= \rho {\cal R}T$, where ${\cal R}$ is the specific gas constant.  A useful
variable is the potential temperature, $\Theta(\bfx,t) \equiv T(P_{\rm
  ref}/p)^\kappa$, where $p_{\rm ref}$ is a constant reference pressure and
$\kappa \equiv {\cal R}/c_p$: when $\dot{q}_{\rm net}\! =\! {\cal D}_T = 0$,
$\Theta$ is materially conserved (i.e. $\D \Theta / \D t = 0$).  The equations
are also supplemented with the `free-slip' boundary condition (i.e. $\D p / \D t
= 0$) at the top and bottom $p$-surfaces; note that the top and bottom
boundaries are material surfaces, across which no mass is transported.  With
these boundary conditions, the equations permit a full range of large-scale
motions for a stably-stratified, un-ionized atmosphere -- with the exception of
sound waves: in equations~(\ref{eq:pe}), sound waves are filtered out from the
full compressible hydrodynamics equations\footnote{Although sound waves are not
  admitted, the primitive equations are still compressible, as $\D\rho / \D t
  \ne 0$.} via the combination of the hydrostatic balance condition, described
by equation~(\ref{eq:pe}b), and the free-slip boundary conditions at the top and
bottom.  However, the results presented in this study also apply to simulations
solving the full Navier-Stokes (non-hydrostatic) equations employing a similar
physical setup, as fast gravity waves admitted by both the hydrostatic and
non-hydrostatic equations approach the speed of sound waves in the modelled
atmospheres (see Table~\ref{tab:params}).

In this work, equations~(\ref{eq:pe}) and (\ref{eq:hyper}) are solved in the
`vorticity-divergence and potential temperature' form\footnote{the curl and
  divergence of equation~(\ref{eq:pe}a), along with equation~(\ref{eq:pe}d) in
  terms of the potential temperature} in the BOB code, a highly-accurate and
well-tested code for extrasolar planet flow applications
\citep{PoliCho12,Polietal14,Choetal15}.  BOB is essentially a multi-layer
extension of the 1-layer codes used in the high-resolution, turbulent studies of
giant planets by \citet{ChoPol96a,ChoPol96b} and \citet{Choetal03,Choetal08}.
The time integration of the equations in all of these codes is performed using a
second-order accurate, leap-frog scheme with a small amount of Robert--Asselin
filter applied to suppress the computational mode arising from the scheme
\citep{Robe66,Asse72}.  The time-step size $\Delta t$ in all the simulations are
such that the Courant-Friedrichs-Lewy (CFL) number \citep[e.g.][]{Stri04,Durr10}
is well below unity -- typically $ < 0.3$.

\begin{table}
  \caption{Physical, Numerical and Scale Parameters:\quad $^{(a)}$~based
    on $c_p$; $^{(b)}$~for $\{{\rm H}_2, {\rm He}\}$; $^{(c)}$ at $p =
    0.1$\,MPa; $^{(d)}$ at $p = 1$\,KPa }
  \label{tab:params}
  \begin{tabular}{llll}
    \hline Planetary rotation rate & $\Omega$ & 2.1$\times$10$^{-5}$ 
    & s$^{-1}$ \\ 
    Planetary radius & $R_p$ & 10$^8$ & m \\ 
    Surface gravity & $g$ & 10 & m\,s$^{-2}$ \\ 
    Specific heat at constant $p$ & $c_p$ & 1.23$\times$10$^4$ 
    & J\,kg$^{-1}$\,K$^{-1}$ \\ 
    Specific gas constant$^{(a,b)}$  & ${\cal R}$ & 3.5$\times$10$^3$ 
    & J\,kg$^{-1}$\,K$^{-1}$ \\ 
    \\ 
    Initial temperature$^{(c)}$ & $T_m$ & 1600 & K \\ 
    `Equil.' sub-stellar temp.$^{(c)}$ & $T_{e_{\rm d}}$ & 1720 & K \\
    `Equil.' anti-stellar temp.$^{(c)}$ & $T_{e_{\rm n}}$ & 1480 & K \\
    Thermal relax. time $^{(d)}$ & $\tau_{\rm th}$ &  $\approx\! 10^5$ & s \\  
    Pressure at top & $p_{\rm top}$ & 0 & MPa \\
    Pressure at bottom & $p_{\rm bot}$ & $[0.1,10]$ & MPa \\     
    Pressure w/o forcing   & $p_0$ & $ \ge 1 $ & MPa \\
    \\
    Truncation wavenumber & T & $[21,682]$ & \\ 
    Number of levels (or layers) & L & $[3,1000]$ & 
    \\ Max. sectoral wavenumber & $M$ & $= \mbox{T}$ &
    \\ Max. total wavenumber & $N$ & $= \mbox{T}$ & \\ 
    Dissipation operator order & $\frakp$ & $[1,8]$ & \\
    Viscosity coefficient & $\nu_{2\frakp}$ & (see text) &
    m$^{2\frakp}$\,s$^{-1}$\\
    (Hyper)dissip. wavenumber & $n_{d(2\frakp)}$ & (see text) & \\ 
    \\
    Vertical length scale & ${\cal H}$ & $\sim\! {\cal R}T_m / g$ & m \\
    Horizontal length scale & ${\cal L}$ & $\!\ga R_p / 20$ & m \\
    Maximum jet speed & ${\cal U}$ & $\!\ga 2\!\times\! 10^3$ & m\,s$^{-1}$ \\
    Sound speed$^{(c)}$ & $c_s$ & $\approx\! 2.8\!\times\! 10^3$ & m\,s$^{-1}$ \\ 
    Dissipation time-scale & $\tau_d$ & $\sim\! 2\!\times\! 10^5$ & s \\ 
    Brunt-V\"ais\"al\"a frequency & ${\cal N}$ & $\sim\! 2\!\times\! 10^{-3}$ 
    & s$^{-1}$ \\ 
    Rossby number & $R_{\rm o}$ & $\equiv {\cal U} / (\Omega{\cal L})$ &  \\ 
    Froude number & $F_{\rm r}$ & $\equiv {\cal U} / \sqrt{g{\cal H}}$ & \\ 
    Rossby deformation scale & ${\cal L}_{\cal R}$ & $\equiv\sqrt{g{\cal H}}/\Omega$ & \\
    \hline
    \end{tabular}
\end{table}

For each $p$-surface, the code transforms the equations to the spectral space
with a `triangular truncation' -- i.e. up to $N\!  =\! M$ wavenumbers retained
in the Legendre expansion,
\begin{equation}
  \xi(\lambda,\mu,t)\ =\ \sum^M_{m = -M} \sum^N_{n = |m|} \xi^m_n(t)\,
  Y^m_n(\mu,\lambda)\, .
\end{equation}
Here $\xi$ is an arbitrary scalar field; $n$ and $m$ are the total and sectoral
wavenumbers, respectively, with $n \in \mathbb{N}\!\equiv\!\{0,1,2,\ldots\}$ and
$m \in \mathbb{Z}\!\equiv\!\{\ldots,-2,-1,0,1,2,\ldots\}$; $(N,M) =
(\max\{n\},\max\{m\})$; $Y^m_n(\lambda,\mu) \equiv P^m_n(\mu)\,e^{i m \lambda}$
are the spherical harmonic functions, where $\mu\!\equiv\!\sin\phi$ and $P^m_n$
are the associated Legendre functions; and, $\xi^m_n(t)$ are the Legendre
coefficients.  The set $\{Y^m_n\}$ are the eigenfunctions of the Laplacian
operator in spherical coordinates:
\begin{equation}
  \grad^2\,Y^m_n\ =\ -\!\left[\frac{n(n + 1)}{R_p^{\ 2}}\right]Y^m_n\, ,
\end{equation}
where
\begin{equation}
  \grad^2\ =\ \frac{1}{R_p^{\ 2}}\left\{\frac{\del}{\del\mu}
    \left[\left(1 - \mu^2 \right)\frac{\del}{\del\mu}\right]\, +\, 
    \frac{1}{1 - \mu^2}\frac{\del^2}{\del\lambda^2}\right\}\, .
\end{equation}
The $\{Y^m_n\}$ constitutes a complete, orthogonal expansion basis
\citep[e.g.][]{ByroFull92}.  Note that, modulo ${\cal C}$,
equation~(\ref{eq:hyper}) reduces to the Laplacian operator acting on $\chi$
when $\frakp = 1$.  Note also that a representation in spectral space with a
truncation wavenumber~T (not to be confused with the temperature~$T$, and
equalling $N\! =\! M$ in triangular truncations) is transformed to a Gaussian
grid in physical space with approximately $(3\mbox{T},3\mbox{T}/2)$ points in
the $(\lambda,\phi)$-direction: a table of grid sizes for different T numbers
are provided for the reader's convenience (Table~\ref{tab:trunc}).  However, the
Gaussian grid\footnote{used to effect transforms of nonlinear products in
  equations~(\ref{eq:pe}) \citep{Orsz70,Eliaetal70} and to aid in dealiasing
  \citep{Orsz71}} should {\it not} be directly compared with the grid of a
finite-difference (or other grid-based) methods, as the Gaussian grid is
effectively equivalent to a much higher resolution than a finite-difference grid
with the same number of points as the former grid.  This is due to the
pseudospectral method's accuracy and convergence properties: a smooth field is
accurately represented by 2 to 3 points on the Gaussian grid, whereas 6 to 10
points are nominally needed on a finite-difference grid
\citep[e.g.][]{Boyd00,Durr10}.  The use of $\frakp \gg 1$ dissipation order in
spectral methods also results in a comparatively much higher effective
resolution, due to the narrower range of dissipated wavenumbers
\cite[e.g.][]{ChoPol96a}.

Vertically, the domain is decomposed into $\mbox{L}\in\mathbb{Z}^+$
uniformly-spaced points (or layers) in the $p$-coordinate.  Along this
direction, a second-order finite-difference scheme is used -- as is common in
codes solving equations~(\ref{eq:pe}) \citep[e.g.][]{Durr10}.  Given the range,
$p \in [p_{\rm top},p_{\rm bot}]$, the dynamically active levels $p_k$ for $k
\in [1,\mbox{L}]$ are located at
\begin{equation}
  p_k\ =\ \Big(k - \frac{1}{2}\Big)\Big[\frac{p_{\rm bot} - p_{\rm top}}{\mbox{
        L}}\Big]\, .
\end{equation}
The $p_{\rm top}$ and $p_{\rm bot}$ surfaces are dynamically {\it not} active,
but they enforce the boundary conditions.  Note that many studies employ a
$\log(p)$-spacing \citep[e.g.][]{LiuShow13,Choetal15}.  However, the difference
in the vertical spacing does not alter the main conclusions presented in this
paper in any qualitative way.

  \begin{table}
    \begin{center}
      \caption{Truncation wavenumber and corresponding Gaussian grid:}
  \label{tab:trunc}
  \begin{tabular}{cc}
    \hline
   \quad  Truncation \quad \quad & \qquad Grid\ \ (longitude $\times$
    latitude) \quad \quad  \\ 
    \hline    
    T682 & $2048 \times 1024$ \\ 
    T341 & $1024 \times 512$  \\ 
    T170 & $512 \times 256$  \\ 
    T85  & $256 \times 128$  \\ 
    T42  & $128 \times 64$  \\
    T21  & $64 \times 32$  \\ 
    \hline
  \end{tabular}
    \end{center}
  \end{table}

Explicit dissipation plays an important role in this work, as have been in
essentially all atmospheric circulation and global climate simulation works
\citep[e.g.][and references therein]{HamiOhfu08,Lauretal11}.  The general
effects of dissipation, including hyperdissipation, on fully-developed
turbulence for the $\mbox{L} = 1$ case is described in detail in
\citet{ChoPol96a}.  As in that work, a rational procedure is used in this work
to estimate the optimal value of $\nu_{2\frakp}$.  We choose $\nu_{2\frakp}$ so
as to damp oscillations near the truncation wavenumber T on an $e$-folding time
scale $\tau_d$\,:
\begin{equation}\label{eq:dissip}
  \nu_{2\frakp}\ \approx\ \frac{1}{\tau_d} \left[\frac{R_p^{\ 2}}{\mbox{T}\,
      (\mbox{T} + 1)}\right]^{\frakp}\, .
\end{equation}
The precise value is chosen heuristically by examining the kinetic energy
spectrum and physical space fields over time, in a series of carefully-prepared
simulations.  It is important to note that, while the procedure is generically
applicable, the precise parameter value and, more broadly, general solution
characteristics (such as `the critical resolution for convergence') is problem
as well as setup specific.  Given the above-mentioned association with the
poorly-understood region of the parameter space, {\it a separate convergence
  test for each problem and setup is strongly advised}, particularly for
extrasolar planet flow studies -- as seen below and noted in \citet{ThraCho11},
\citet{Polietal14}, and \citet{Choetal15}.

The primary goal of the present study is to assess rigorously the numerical
convergence of current extrasolar planet atmospheric flow simulations {\it with
  a setup that is commonly-employed} \citep[e.g.][]{LiuShow13,Choetal15}.  The
chosen setup is for a model hot-Jupiter, HD209458b (Table~\ref{tab:params}).  As
in many studies, the thermal forcing -- the $\dot{q}_{\rm{\tiny net}} / c_p$
term in equation~(\ref{eq:pe}d) -- is crudely represented by a simple, linear
relaxation on a timescale of $\tau_{\rm th}(p)$ to a specified `equilibrium'
temperature distribution $T_e(\lambda,\phi,p)$; note that here both $\tau_{\rm
  th}$ and $T_e$ are independent of the flow.  Details of the $\tau_{\rm th}$
and $T_e$, as well as the initial $T$ are given in \citet{LiuShow13} and
\citet{Choetal15}.  Note also that, unlike in \citet{LiuShow13}, a strong
Rayleigh dissipation in equation~(\ref{eq:pe}a) is not applied near the bottom
of the domain in the present work.  As reported by \citet{Choetal15}, such a
dissipation coerces the flow to a dynamically simple state -- one essentially
devoid of vortices and waves.  This is very different than the states reached by
all the simulations in this paper.  Because such dissipation is physically
arguable for giant planets and because employing additional dissipation or
energy-conserving schemes (as a strategy to preserve stability, for example) can
greatly distort solutions \citep{Boyd00}, we purposely avoid this expediency in
order to provide a more lucid account.

From hereon, {\it the planetary radius $R_p$ and rotation period $\tau$
  ($\equiv~2\pi/\Omega = 3.025\!\times\!  10^5$\,{\rm s}) are used as the length
  and time scales}, respectively -- whenever clarity is not at risk.  For
example, $\nu_{2\frakp}$ is in the units of $R_p^{2\frakp}\,\tau^{-1}$, and $n$
is in the units of $R_p^{-1}$; however, {\it the temperature $T$ and pressure $p$
  remain in the units of {\rm K} and {\rm MPa}}, respectively.  Then, for all
the simulations discussed in this paper, $p_{\rm top} = 0$, but $p_{\rm bot}$ is
either $ =\! 0.1$ or $ >\! 0.1$.  If $p_{\rm bot} = 0.1$, the modelled
atmosphere is designated, a `shallow atmosphere'; else, it is designated, a
`deep atmosphere'.  In this study, solutions are compared in the spectral,
physical and temporal spaces.  Examinations in all three spaces are required for
a robust assessment of convergence, as it can at times appear to be attained in
one or even two of the spaces.

\section{Results}\label{sec:results}

\begin{figure}
  \centerline{\includegraphics[scale=.11]{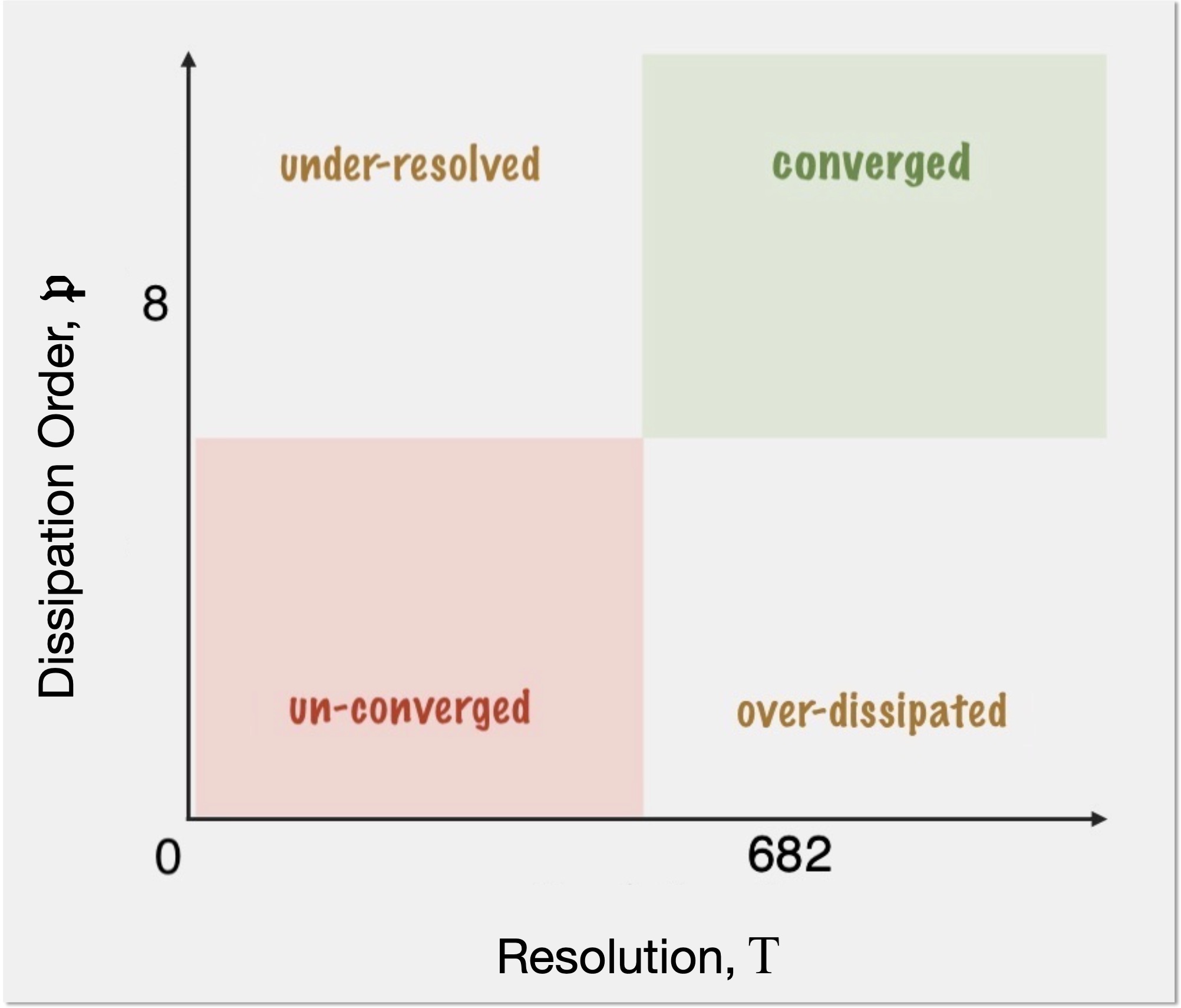}}
  \caption{Summary of convergence -- in dissipation order--horizontal resolution
    ($\frakp$--T) space for simulations with $[0.0,0.1]$\,MPa vertical domain.
    High T {\it and} high $\frakp$ are required for convergence --
    e.g. T~$\ge$~341 {\it and} $\frakp > 6$ (green shaded area).  Simulations
    with low T erroneously `converge' to an unphysical state because they do not
    include interactions with small-scales.  Simulations with low $\frakp$ also
    erroneously `converge' to an unphysical state, as energy is effectively
    dissipated at {\it all} scales (including the large-scales).  Simulations
    with inadequate T {\it and} low order $\frakp$ are not converged (red shaded
    area) because they are both under-resolved and over-dissipated.  A high
    vertical resolution (e.g. L $\ga\! 10$, over the $p$-range given above) --
    along with high T and high $\frakp$ -- is also necessary for convergence.
    These requirements for convergence are primarily due to the physical nature
    of the planet (e.g. relatively high $T_m$ and moderate $\Omega$, common
    among many short-period objects), and the simulation setup employed -- in
    particular, high-amplitude thermal forcing with a short relaxation time,
    leading to a high-speed turbulent flow.}
  \label{fig:converged}
\end{figure}

Fig.~\ref{fig:converged} summarizes the basic result of this study.  It
illustrates the `area of convergence' in $\frakp$--T (order--resolution)
parameter space.  As depicted, {\it high {\rm T} and high $\frakp$ are both
  required for convergence} (green shaded area).  Even with a high~T,
simulations with a low~$\frakp$ `converge' to an erroneous state, since energy
is in effect dissipated at all scales -- including the large-scales.  With a
high~$\frakp$ but a low~T, simulations also `converge' to an unphysical state,
as they preclude small-scales from interacting (since the scales are not
represented).  With both low~T and low~$\frakp$, simulations are not converged
(red shaded area), as they are both under-resolved {\it and} over-dissipated.
The lack of convergence and accuracy outside the green shaded area is
principally caused by truncation and discretization errors, as well as
ill-effects from `stability-enhancing' strategies.  We note here that most of
the past extrasolar planet atmosphere simulations reside in the red shaded area
\citep[see discussions
  in][]{ThraCho11,PoliCho12,Polietal14,Choetal15,Choetal19}.

In an extended parameter space which includes the vertical resolution, nominally
$\mbox{L} \approx 10$ (over the $p \in [0.0,0.1]$ range) is also necessary for
convergence -- along with the high~T and high~$\frakp$ already discussed.  For
larger $p$-range, larger L is necessary.  This is related to the amplitude and
type of forcing applied (prescribed with $\max\{|\grad\!_p T|\} \approx 10^3$\,K
on an extremely short relaxation time $\tau_{\rm th}$) -- and, crucially, the
complex turbulent flow resulting from it.  The generated flow contains
large-scale meandering high-speed jets\footnote{In fact, the jet near the
  equator is nearly always supersonic (${\cal U} / c_s~>~1$).  However,
  supersonic flows are not physically valid for equations~(\ref{eq:pe}) and
  free-slip boundary conditions \citep{Choetal15}.} and large-scale vortices
(generally in pairs), along with many energetic small-scale vortices and waves
that strongly influence the large-scale flow.  Note that a physically more
sophisticated forcing, based on coupling with a one-dimensional radiative
transfer model \citep[e.g.][]{Showetal09}, does not mitigate the L requirement
-- as well as the T and $\frakp$ requirements -- because similar flows are still
generated.  As emphasized in \citet{Choetal15}, the {\it a}geostrophic nature of
the modelled atmosphere (i.e. Rossby number $R_{\rm o}$ and Froude number
$F_{\rm r}$ both of order unity) and the forcing and initialization setup
commonly used all work in concert to impose an uncommonly stringent requirement
on numerical codes.

In what follows, we first present the results from a very high resolution
simulation -- as a fiducial reference solution.  Then, we discuss the
convergence behaviour with respect to the horizontal resolution T, dissipation
order $\frakp$ and coefficient $\nu_{2\frakp}$ and vertical domain range
(shallow/deep atmosphere) and resolution L.  Always equatable solutions within a
set of simulations are compared, with each other (as well as with the reference
solution).  Here `equatable' refers to that quality shared by simulations for
which the value of a single parameter is different while those for all others
are identical -- possibly with the exception of a parameter or two that must be
adjusted concomitantly to maintain constancy of certain `global' property
(e.g. $\Delta t$ for the CFL number).  The procedure for adjusting
$\nu_{2\frakp}$ with varying T or $\frakp$, described in
section~\ref{sec:method} and used in virtually all the simulations discussed in
this paper, is another example of rendering simulations equatable -- i.e. as
$\frakp$ is varied, $\nu_{2\frakp}$ is changed to maintain the same dissipation
rate at the truncation wavenumber, $n = \mbox{T}$.  Through the equatable
comparisons, we find that the $\{\mbox{T},\frakp,\mbox{L}\}$ requirement for
convergence is robust up to the highest horizontal resolution investigated in
this study (T682), and it is expected to hold at higher resolutions valid for
equations~(\ref{eq:pe}).

\subsection{Reference Solution}

\begin{figure*}
  \centerline{\includegraphics[scale=.64]{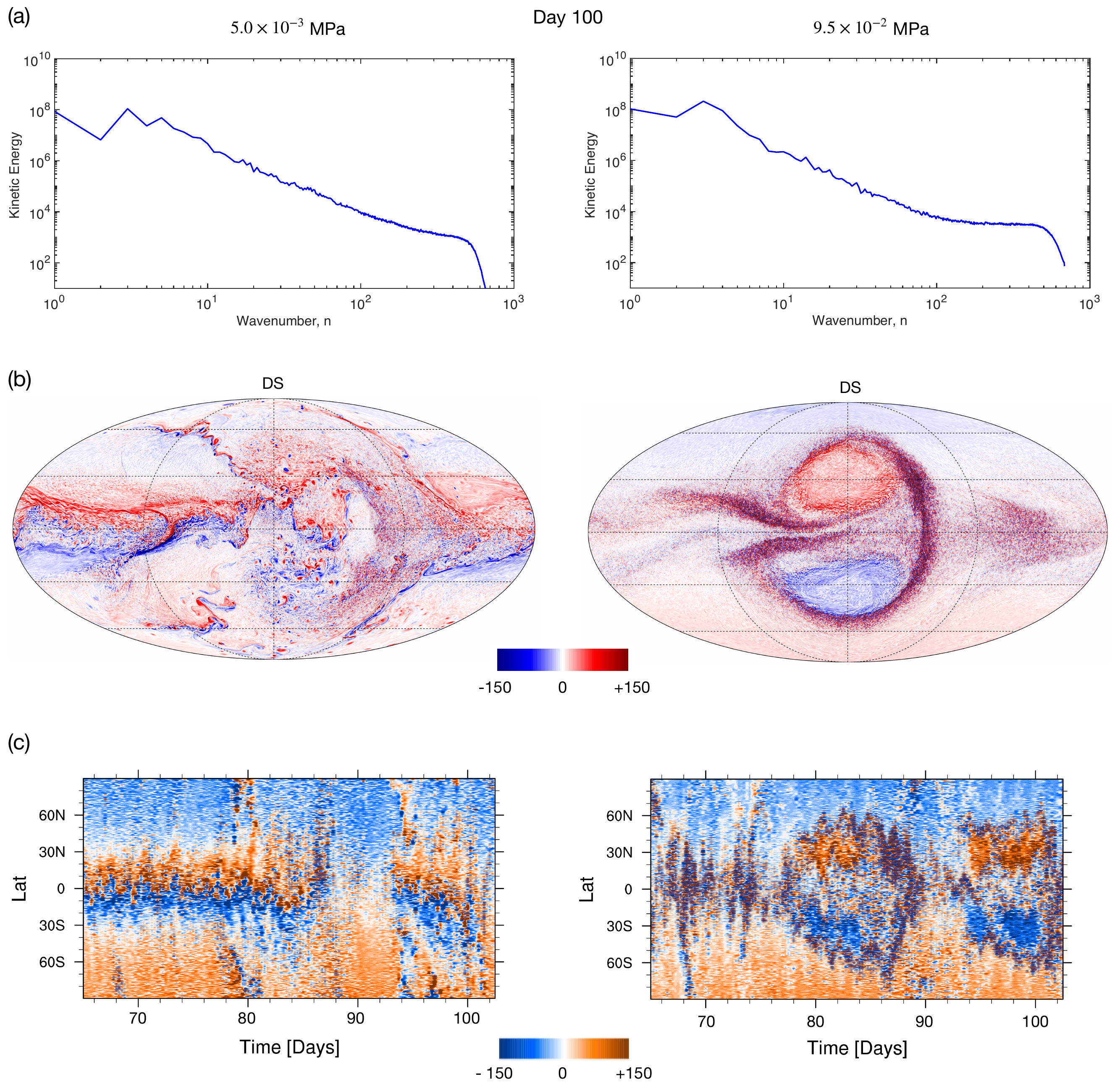}}
  \caption{The reference solution -- from a T682L20 simulation at $p$-levels,
    $5.0 \times 10^{-3}$ (left column) and $9.5\times 10^{-2}$ (right column),
    (in units of MPa): (a)~kinetic energy spectra at $t = 100$ (in units of the
    planetary rotation period $\tau$); (b) relative vorticity fields (in units
    of $\tau^{-1}$) in Mollweide projection centred on the sub-stellar point;
    (c) $\phi$--$t$ Hovm\"{o}ller plots of the relative vorticity at the
    longitude, $\lambda\!  =\!  0$, for $t \in [65, 102.5]$ duration.  The
    dissipation order is $\frakp\! =\! 8$, viscosity coefficient $\nu_{16} =
    2.3\!\times\!  10^{-48}$ (in units of $R_p^{16}\,\tau^{-1}$, where $R _p$ is
    the planetary radius) and time-step size is $\Delta t = 2\!\times\!  10^{-5}$
    (in units of $\tau$).  After an initial build-up period, the spectra at all
    $p$-levels are broad and show a shallow sub-spectra for wavenumbers (in units
    of $R_p^{-1}$), $n \ga 100$ (particularly noticeable at $p \ga
    10^{-2}$)~(a), reflecting the flows~(b).  At $p = 5.0 \times 10^{-3}$, a
    high-speed, zonally-{\it a}symmetric equatorial jet is unstable and
    generates a large number of medium- and small-scale, long-lived vortices (b,
    left); the fast eastward jet is periodically countered by a slower eastward
    jet -- leading to the former's disruption, for example, at $t \approx 86$
    (c, left).  At $9.5\times 10^{-2}$, the flow is dominated by two modons
    straddling the equator, a cyclonic pair (demarcated by a high-speed, sharp
    front surrounded by a large number of small-scale vortices) and a much
    weaker, diffused anticyclonic pair covering a significant portion of the
    remaining space (b, right); the cyclonic modon undergoes periodic life-cycles
    (c, right).}
  \label{fig:refsol}
\end{figure*}

In Fig.~\ref{fig:refsol}, we present the reference solution, which is at T682L20
resolution with $\frakp\!  =\! 8$ hyper-viscosity (i.e. $\Del^{\rm 16}$
dissipation operator).  As listed in Table~2, the size of the dealiasing,
Gaussian grid in physical space used here is $2048\!\times\!  1024$ points
\citep[see e.g.][and references therein for discussions of dealiasing and
  Gaussian grid]{ChoPol96a,Boyd00,ThraCho11}.  The (non-dimensionalized)
hyperdissipation coefficient and time-step size are $\nu_{16} = 2.3\!\times\!
10^{-48}$ and $\Delta t = 2\!\times\! 10^{-5}$, respectively.  The vertical
range of the simulation domain is $p \in [0.0,0.1]$ (in units of MPa) -- i.e. of
a shallow atmosphere.  The figure illustrates the three types of diagnostics
used to assess convergence, as discussed in section~\ref{sec:method}: the
kinetic energy spectra, instantaneous flow fields and temporal evolution of the
flow fields. Each diagnostic is applied to the levels near the top (left column)
and bottom (right column) of the domain.

Fig.~\ref{fig:refsol}a shows the kinetic energy spectrum $\hat{\cal E}(n)$ of
the flow field on day, $t\! =\! 100$, of the simulation at the indicated
$p$-levels.  Here $\hat{\cal E}(n)$ is an average along the $m$-direction in
spectral space for each wavenumber $n$.  The figure illustrates several generic
features of $\hat{\cal E}$, present at all $p$-levels in a shallow atmosphere:
1)~$\hat{\cal E}$ is broad, due to the fact that a large fraction of high
wavenumbers contain a significant amount of energy; 2)~much more energy is
contained in the low wavenumbers than in the high wavenumbers, consistent with
the strong stratification of the modelled atmosphere (signified by ${\cal N}^2 /
\Omega^2 \gg 1$, where ${\cal N}$ is the Brunt-V\"{a}is\"{a}l\"{a} frequency);
and, 3)~a very shallow sub-spectrum (for $n \ga 100$) exists -- particularly
noticeable at most of the $p$-levels other than at a narrow range of levels near
the top (cf. right and left panels).  In contrast, the sub-spectrum for low
wavenumbers possesses a much steeper slope.  Feature 3) is a consequence of the
{\it a}geostrophy of the modelled atmosphere: in {\it a}geostrophic atmospheres,
small-scale vortices and gravity waves (`eddies') are readily generated and
persist over long time.  Here all of the above features already clearly
demonstrate the need for high resolution ({\it much higher than currently
  typical}) in extrasolar planet atmospheric flow simulations.

Fig.~\ref{fig:refsol}b shows the flow fields from which the corresponding
spectra in Fig.~\ref{fig:refsol}a have been obtained.  In
Fig.~\ref{fig:refsol}b, the instantaneous relative vorticity field
$\zeta(\lambda,\phi)$ (in the units of $\tau^{-1}$) is shown in Mollweide
projection, centred on the sub-stellar point: $(\lambda,\phi)\!  =\!  (0,0)$.
Note that the modelled planet is assumed to be in a 1:1 spin--orbit synchronized
state.  Here $\zeta > 0$ (red) in the northern hemisphere and $\zeta < 0$ (blue)
in the southern hemisphere indicate local rotation in the same sense as
$\bm{\mathit{\Omega}}$ -- and vice versa.  All flow fields in this paper are
shown in the same projection, and centred at the sub-stellar point, so that the
entire flow field can be seen and compared easily with these fields (as well as
with each other).  In general, the flow near the top (left) is distinct from the
flow in the rest of the domain (e.g. right), consistent with the spectral
behaviour (Fig.~\ref{fig:refsol}a).  Overall, the flow is strongly barotropic
(vertically aligned) and only weakly baroclinic (vertically slanted) because of
the flow near the top -- as reported by \citet{ThraCho10}, in an earlier study
using a different code.  Broadly, the flow field at all $p$-levels is
characterized by an undulating equatorial jet and curved, planetary-scale fronts
in the northern and southern hemispheres.

More specifically, the equatorial flows in Fig.~\ref{fig:refsol}b are flanked by
two planetary-scale modons\footnote{A `modon' is a stable vortex-pair structure,
  in which the two vortices have opposite signs -- i.e. a vortical dipole
  \citep{Ster75}.} -- a pair of cyclonic\footnote{Cyclonicity is defined by the
  sign of $\bm{\mathit{\zeta}} \cdot \bm{\mathit{\Omega}}$; for a cyclone it is
  positive, and for an anticyclone it is negative.}  vortices on the day side
and a pair of anticyclonic vortices on the night side.  The modons begin to form
at the start of the simulation, with their centroids at the equator a distance
(in units of $R_p$) of $\sim\!\pi$ apart in longitude at `full maturity'.  As
early as $t \approx 2$, they start to lose their north--south symmetry, showing
differentiation in the $\max\{|\zeta|\}$ as well as the areal extent (despite
the north--south symmetry of the prescribed forcing).  This occurs because
modons are not formal solutions to equations~(\ref{eq:pe}), with or without the
forcing, and they radiate Rossby and gravity waves -- even in the
quasi-geostrophic regime at the beginning of the simulation (when the flow speed
is low).  Also apparent is the symmetry breaking between the two modons: the
cyclonic modon is much stronger than the anticyclonic modon; this east--west
symmetry is broken from the start of the simulation as the forcing drives
westward propagating Rossby waves.  The intense cyclonic modon (right) emits
large-amplitude gravity waves and induce thousands of small-scale vortices to
form at its periphery, with larger number of cyclones (as well as to trail it).
These vortices preferentially strengthen the positively-signed,
northern-hemisphere half of the cyclonic modon and screen the negatively-signed,
southern-hemisphere half of the cyclonic modon -- thus contributing to the
north--south asymmetry; significantly, we note that the symmetry is broken only
at (and above) T341 resolution.  In contrast, its high-latitude position
notwithstanding, the anticyclonic modon is barely visible at both $p$-levels
because of the much more diffused interiors and the lack of bounding fronts.  It
is important to understand that, when converged, all of the above features are
generic in all simulations employing the aforementioned setup.

Additionally, the cyclonic modon is very dynamic and executes a complex set of
motions.  Typically, the modon first forms just to the west of the sub-stellar
point and initially moves in the westward direction.  It then reverses direction
and starts to move eastward, back towards the sub-stellar point.  Near the
sub-stellar point, it `pauses' for a long period (up to $\sim$10 days).  Then,
it heads back in the westward direction, this time migrating past the western
terminator and fully traversing the night side.  Finally, the modon (often)
dissipates completely near the eastern terminator.  Subsequently, the entire
`erratic' motion -- from formation to dissipation -- repeats quasi-periodically
over the duration of the simulation.  Throughout this peregrination, a large
number of small-scale vortices are generated and {\it vigorous mixing on the
  planetary-scale is induced, which crucially affect the temperature as well as
  radiatively- and chemically-active species distributions}.  During the
`eastward-migration phase', if the modon manages to migrate past the sub-stellar
point (instead of turning back) and reaches close to the eastern terminator, the
constituent cyclones spread apart and sometimes even decouple, dissolving the
modon.  The overall motion described above is in sharp contrast to the
essentially steady, simple westward translation observed in low-resolution
simulations with low~$\frakp$, as will be seen in the ensuing sections.  The
complex motion is due to the nonlinear interactions with energetic small-scale
eddies, not present in low~T and/or low~$\frakp$ simulations.

\begin{figure*}
 \centerline{\includegraphics[scale=.48]{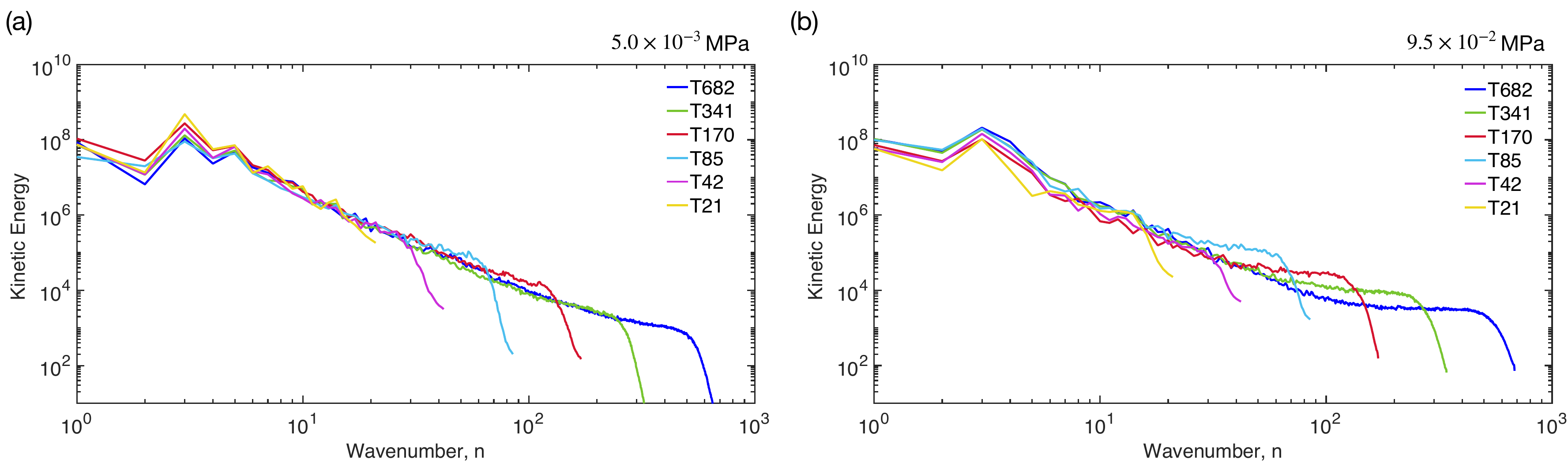}}
 \caption{Kinetic energy spectra at $t = 100$ from simulations which are
   identical to the simulation in Fig.~\ref{fig:refsol} -- except the horizontal
   resolution~T, viscosity coefficient~$\nu_{16}$, and the time-step size
   $\Delta t$.  The vertical resolution of all the simulations presented is the
   same ($\mbox{L}\!  =\!  20$); but, because T varies, $\Delta t$ is halved for
   each doubling of the resolution -- to ensure a uniform CFL condition
   \citep{Boyd00}, given a fixed maximum flow speed.  The $\frakp\!  =\! 8$
   dissipation operator is used in all the simulations shown: for $(\mbox{T}21,
   \mbox{T}42, \mbox{T}85, \mbox{T}170, \mbox{T}341, \mbox{T}682)$ resolutions,
   $\nu_{16}$ values are correspondingly $(2.7 \times 10^{-24}, 4.2 \times
   10^{-29}, 6.4 \times 10 ^{-34}, 9.8 \times 10^{-39}, 1.5 \times 10^{-43}, 2.3
   \times 10^{-48})$.  The $p$-levels in (a) and (b) are as in
   figure~\ref{fig:refsol}; and, the T682 spectra in (a) and (b) are reproduced
   from figure~\ref{fig:refsol}a.  This figure shows that simulations are
   definitely not converged below the T341 resolution.  Spectrally, convergence
   ``appears'' to be achieved at $p = 5.0 \times 10^{-3}$ with T341 resolution,
   at least up to $n \la 200$; however, the T341 and T682 simulations still
   behave somewhat differently in physical space -- as is suggested by the
   differences in the spectra for $n \ga 40$ at $p = 9.5 \times 10^{-2}$~(b).
   Full (i.e. spectral, physical, {\em and} temporal) convergence in the
   qualitative sense is achieved only at T341 resolution and only at the top of
   the atmosphere, for the physical setup employed.  Note that T341 resolution
   here corresponds to roughly a minimum of $2048\!\times\!1024$
   finite-difference grid resolution.}
  \label{fig:specrez}
\end{figure*}

The complex time-dependent behaviour can be discerned in Fig.~\ref{fig:refsol}c,
which shows the $\phi$--$t$ Hovm\"{o}ller plots of the $\zeta$ field at $\lambda
= 0$ for the duration $t \in [65, 102.5]$.  The duration includes the time of
the Figs.~\ref{fig:refsol}a and~\ref{fig:refsol}b.  The plots in
Fig.~\ref{fig:refsol}c illustrate the main features of the `flow evolution' of a
shallow atmosphere simulation at high resolution.  Near the top of the
domain~(left), a high-speed eastward jet\footnote{identified by the zonally-{\it
    a}symmetric `band' of $\zeta$ with a `jump' across the equator and bounded
  by sharp gradients at the band's edges; see, in particular, the night side
  (N.B. jets are present at both $p$-levels).} is obstructed by a much slower
eastward jet at the equator, leading to the former's disruption just east of the
sub-stellar point at $t \approx 86$: concurrently, planetary-scale curved fronts
break and roll up into vortices (Fig.~\ref{fig:refsol}b, left).  The large
variance of $\zeta$ in the $\phi$-direction is the signature of the congestion
and breaking, clearly identifiable in the plot; and, it is paradigmatic of
numerous such episodes occurring throughout the entire duration of the simulation
(e.g.  $t \in \{78,80,94\}$).  Near the bottom of the domain (right), the
cyclonic modon's dominance (Fig.~\ref{fig:refsol}b, right) in the evolution can
be seen clearly -- including the signature of the aforementioned quasi-periodic
life-cycles.  For example, the following `phases' of the cycle are detected in
the plot on the right: the modon moving westward ($72 \la t \la 76$), reversing
direction and pausing near the sub-stellar point ($77 \la t \la 86$), migrating
westward all the way around the night side ($87 \la t \la 94$), dissipating
completely near the western terminator ($t \approx 94$) and returning to a
quasi-stationary state after forming again just to the west of the sub-stellar
point ($ 94 \la t \la 101$).

\subsection{Numerical Resolution}\label{sec:rez}

In Fig.~\ref{fig:specrez}, we present the kinetic energy spectrum $\hat{\cal
  E}(n)$, taken from simulations which are identical to the simulation of
Fig.~\ref{fig:refsol} in all respects -- except for T (and the correspondingly
adjusted $\Delta t$ and $\nu_{16}$).  As discussed above, $\Delta t$ is halved
for each doubling of the T value to ensure a uniform CFL condition in all the
simulations presented, given the approximately constant maximum flow speed
${\cal U}$ ($\approx\! 12.25$, in the units of $R_p\,\tau^{-1}$) across the
simulations; and, $\nu_{16}$ is adjusted so that the energy dissipation rate at
a given wavenumber (e.g. $n = 21$) is same in all the simulations.  In general,
$\hat{\cal E}$ depends on $t$ -- even after reaching equilibration, as the
equilibration state itself can fluctuate quasi-periodically on varying
time-scales for different simulations; in addition, $\hat{\cal E}$ depends on
$p$ as well, as will be seen below.  In the figure, all the simulations shown
have reached equilibration at {\it all} the $p$-levels by $t \!\sim\! 10$; and,
at the time shown ($t = 100$), all the spectra are representative of the mean
equilibration state.  The $p$-levels shown are as in Fig.~\ref{fig:refsol}, and
the reference spectra from Fig.~\ref{fig:refsol}a is reproduced in
Fig.~\ref{fig:specrez} for ease of comparison.

\begin{figure*}
  \vspace*{1cm} \centerline{\includegraphics[scale=.51]{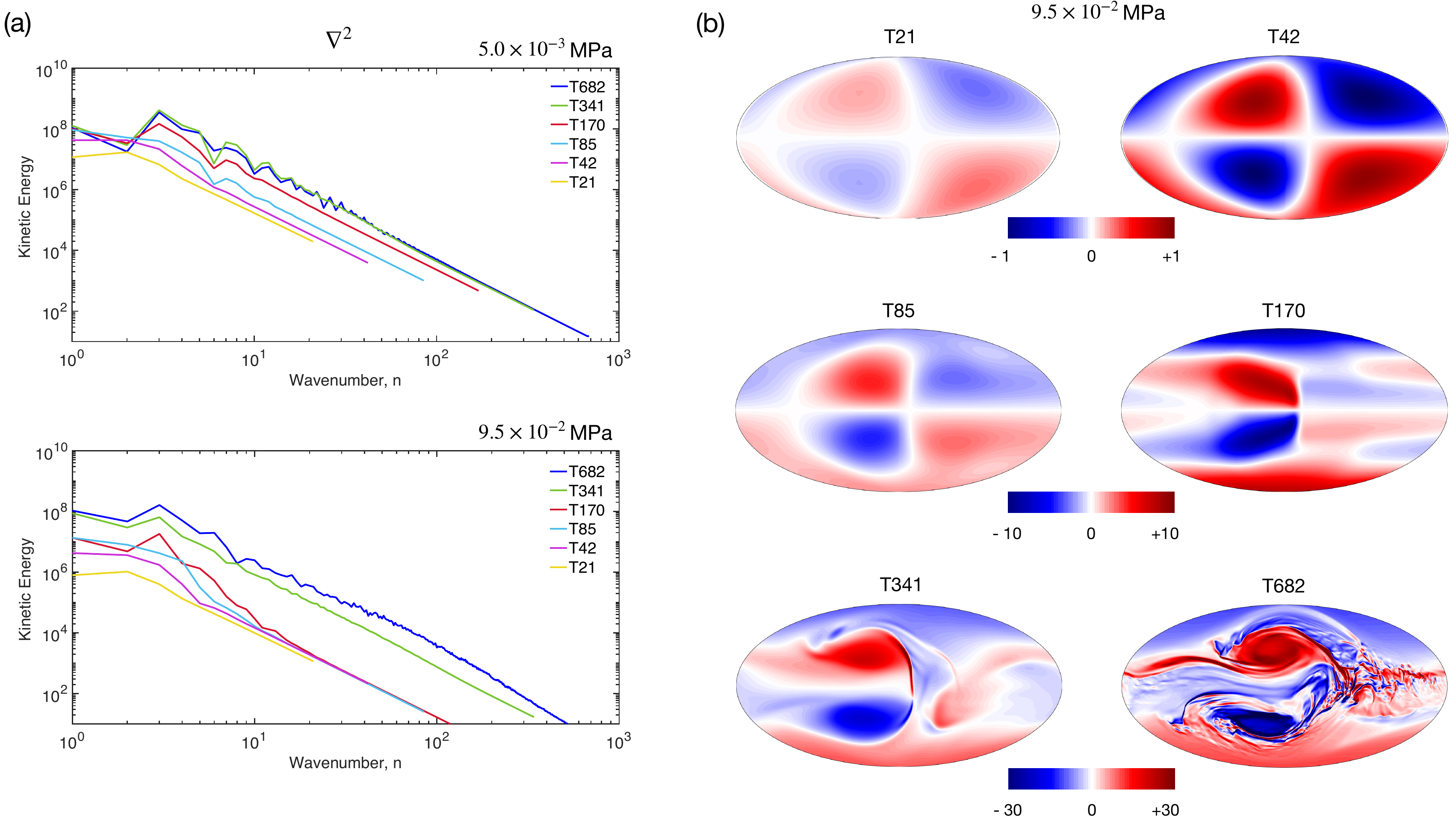}}
  \caption{$\frakp\! =\! 1$ simulations identical in all respects, except the
    horizontal resolution (and accordingly adjusted viscosity coefficient
    $\nu_2$ and time-step size~$\Delta t$): kinetic energy spectra (a) and the
    corresponding vorticity fields (b) at $t = 82$, with $p$-levels as
    indicated.  The $\frakp\! =\! 1$ simulations are not converged until the
    T341 resolution at the top and not at all below, as in the $\frakp\! =\!  8$
    case (cf. Fig.~\ref{fig:specrez}).  Significantly, the dissipation appears
    to affect the entire range of wavenumbers, down to the low wavenumbers --
    even down to $n = 2$ at the T341 resolution for $p = 9.5\!\times\! 10^{-2}$.
    The relative vorticity fields below the T341 resolution exhibit meridional
    (north--south) symmetry (modulo the sign change due to sphericity),
    consistent with the dominance of viscosity seen in (a).  Below the T85
    resolution, the flow structures are diffused and essentially static.  Above
    the T85 resolution, the modon exhibits dynamism that increases with
    resolution.  Sharp, elongated fronts that roll up into small scale features
    emerge only at the T682 resolution; however, the flow is still distinct from
    the $\frakp\!  =\!  8$ case (cf. Fig.~\ref{fig:refsol}b).  The overall
    behavior strongly motivates the use of a higher order dissipation
    (preferrably $\frakp \gg 1$; see Figs.~\ref{fig:T682_del2del16} and
    \ref{fig:delcomp}).}
  \label{fig:del2rez}
\end{figure*}

Fig.~\ref{fig:specrez} shows that simulations are definitely not converged below
T341 resolution.  It also shows the importance of widening the focus to more
than just one $p$-level or vertical region \citep[][]{Choetal15,Choetal19}, as
{\it $\hat{\cal E}$ at different $p$-levels can have different convergence
  properties}.  Consider the $p = 0.005$ level (Fig.~\ref{fig:specrez}a), for
example.  At first glance, convergence appears to be achieved at T42 resolution;
but, spectra at the $p = 0.095$ level (Fig.~\ref{fig:specrez}b) clearly show
that convergence is actually not achieved even at T341 resolution.  In
Fig.~\ref{fig:specrez}b, spectral blocking (the raised `backward-facing step' in
the mid-$n$ region of $\hat{\cal E}$) is apparent in the T85, T170 and T341
spectra: such a spectral feature is produced by aliasing error in simulations
which under-resolve the flow \citep{Boyd00,ThraCho11}.  On closer inspection,
the T85 and T170 (and possibly the T42) spectra at the $p = 0.005$ level also
display weak spectral blocking.  Typically, spectral blocking is more easily
noticed occurring near $n = \mbox{T}$, particularly under quasi-geostrophic
conditions, because of the steepness of $\hat{\cal E}$ (low energy content) in
that region.  Here high wavenumbers (i.e. $n > \mbox{T}$), energized by the
flow, are aliased onto the mid-$n$ region.  The feature also frequently shows up
shortly before a simulation `blows up'.  At lower resolutions, spectral blocking
is altogether, or nearly so, masked by over-dissipation (T21 and T42 spectra).
In this case, crucial information about physics (e.g.  frontal dynamics at the
modon's periphery) is suppressed -- in peril of the simulation's verisimilitude,
as shown below.

Considering spectra at both $p$-levels together, one might be tempted to argue
for convergence predicated on a sub-range of wavenumbers (e.g. convergence at
T85 resolution, for $n \la 20$).  However, the behaviour in physical space of
simulations with up to T341 resolution is still qualitatively different compared
to the behaviour of simulations with T682 resolution -- even at the large
scales.  This is not surprising, given the large difference in $\hat{\cal E}$
for $n \ga 40$ (i.e. most of the $\{n\}$ included in the simulation) and the
non-linearity of equations~(\ref{eq:pe}).  Moreover, {\it because spectral
  blocking is a manifestation of accumulated errors infecting all $n$,
  large-scale behaviours are unreliable in simulations that evince it}.
Clearly, $\hat{\cal E}$ is very useful as a first diagnostic -- especially at
very high resolutions, when {\it plotting} one snapshot of the flow field can
sometimes take more than an Earth day.  In sum, convergence is achieved only at
T341 resolution for the physical setup employed -- and that only at the top
region of the domain.  Note that T341 resolution here corresponds to roughly a
minimum of $2048\!\times\!1024$ finite-difference grid resolution.  As far as we
are aware, past simulations that use the same, or similar, setup have been
performed with much lower resolutions and dissipation orders \citep[e.g.][and
  references therein]{Meno20}.

At this point, the reader may wonder if high-order dissipation is necessary --
or even proper.  After all, the Navier--Stokes equations (from which the
primitive equation derives) are with $\frakp = 1$ dissipation.  We briefly
address this issue here and leave the more detailed discussion for the next
sub-section.  Fig.~\ref{fig:del2rez} demonstrates clearly why the $\frakp = 1$
dissipation is not adequate (up to the resolution presented, for the setup
employed).  In the figure, simulations at different resolutions are presented
with resolution-adjusted $\nu_2$ and $\Delta t$, but otherwise identical.  Here
$t = 82$ for all the simulations.  The time chosen is well after the kinetic
energy time series have reached their equilibrated (i.e. quasi-stationary)
states and remain qualitatively unchanged for up to 300~days.  The overall
behaviour in both the spectral (Fig.~\ref{fig:del2rez}a) and physical
(Fig.~\ref{fig:del2rez}b) spaces strongly suggests the use of $\frakp > 1$ to
prevent over-dissipation (in fact $\frakp > 4$, as demonstrated explicitly in
the next sub-section).

\begin{figure*}
  \centerline{\includegraphics[scale=0.14]{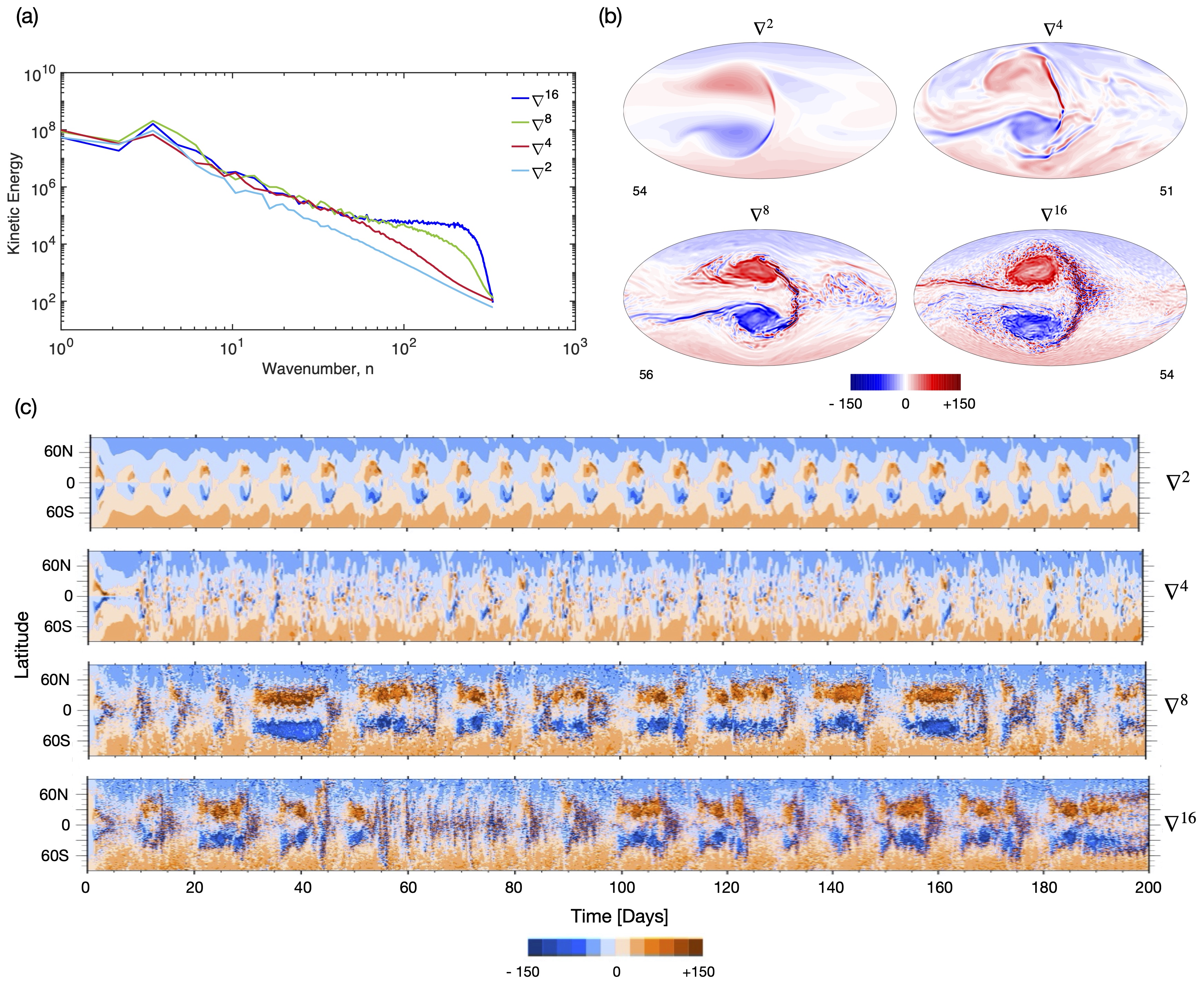}}
  \caption{Kinetic energy spectra (a) of the flow fields (b) at $p = 9.5\times
    10^{-2}$ ($t$ is slightly different in the different frames, chosen from
    $[51,56]$, so that the modons are `in phase') and the $\phi$--$t$
    Hovm\"{o}ller plots of $\zeta$ at $\lambda = 0$ (c) from four T341L20
    simulations -- identical in setup, except for the $\frakp$ in
    $\grad^{2\frakp}$ (and $\nu_{2\frakp}$).  For $\frakp \in \{1,2,4,8\}$,
    $(\nu_2, \nu_4, \nu_8, \nu_{16}) = (10^{-3}, 10^{-8}, 10^{-18},1 0^{-43})$,
    so that the dissipation rate for $n = 341$ is same for all $\frakp$ and
    equatable comparisons can be made between the simulations.  In general,
    lower $\frakp$ dissipates over a broader range of $n$ as well as dissipates
    the higher $n$ more strongly (a).  This is consistent with the more
    diffused, sluggish flow structures seen in the simulations (b).  Note that,
    while these modons are associated with areas of higher temperature compared
    to their surroundings for all $\frakp$ \citep{SkinCho20}, they should {\it
      not} be confused with the `eastward-shifted hot area' at the lower
    $p$-levels (higher altitudes) in lower resolution simulations, oft discussed
    in the literature \citep[see][for more information]{Choetal19}.  There is a
    noticeable difference in the general character of the flow with $\frakp$.
    For example, note the `phase variation' in the modon's position ($t$ is
    given at the bottom in each frame) across the simulations.  This is because
    modons quasi-periodically break up into small-scale storms and subsequently
    reform when $2\frakp \ge 8$, while modons execute a nearly-steady
    translation with minor perturbation when $2\frakp < 8$.  The former pair of
    simulations show $\sim$14~life-cycles of a modon forming on the day-side and
    traversing around the planet with an average period of roughly
    $\sim$12\,$(\pm 5)$~days.}
 \label{fig:delcomp}
\end{figure*}

Fig.~\ref{fig:del2rez}a shows that simulations with $\frakp = 1$ are not
converged, up to T341 resolution (and, in actuality, up to T682 over most of the
domain).  We note here that, significantly, the dissipation affects essentially
the entire range of $n$ -- all the way down to $n = 1$ in nearly all the
simulations presented (cf. Fig.~\ref{fig:specrez}).  Fig.~\ref{fig:del2rez}b
presents the $\zeta$-fields at $p = 9.5\times 10^{-2}$, from which the spectra
in Fig.~\ref{fig:del2rez}a are obtained.  Here simulations with T21, T42 and T85
resolutions display meridional (north--south) symmetry -- indicating the
dominance of explicit dissipation, which is meridionally symmetric.  Below T85
resolution, the flow structures are diffused and essentially static.  At T85 and
above resolutions, a moving cyclonic modon forms westward of the sub-stellar
point.  Crucially, the modon exhibits significant dynamism only at, and above,
the T341 resolution with high~$\frakp$ (cf. Figs.~\ref{fig:refsol}b and
\ref{fig:delcomp}b).  These features are robust, and persist up to long
integration times (e.g. here $t = 300$).

\begin{figure*}
  \centerline{\includegraphics[scale=0.115]{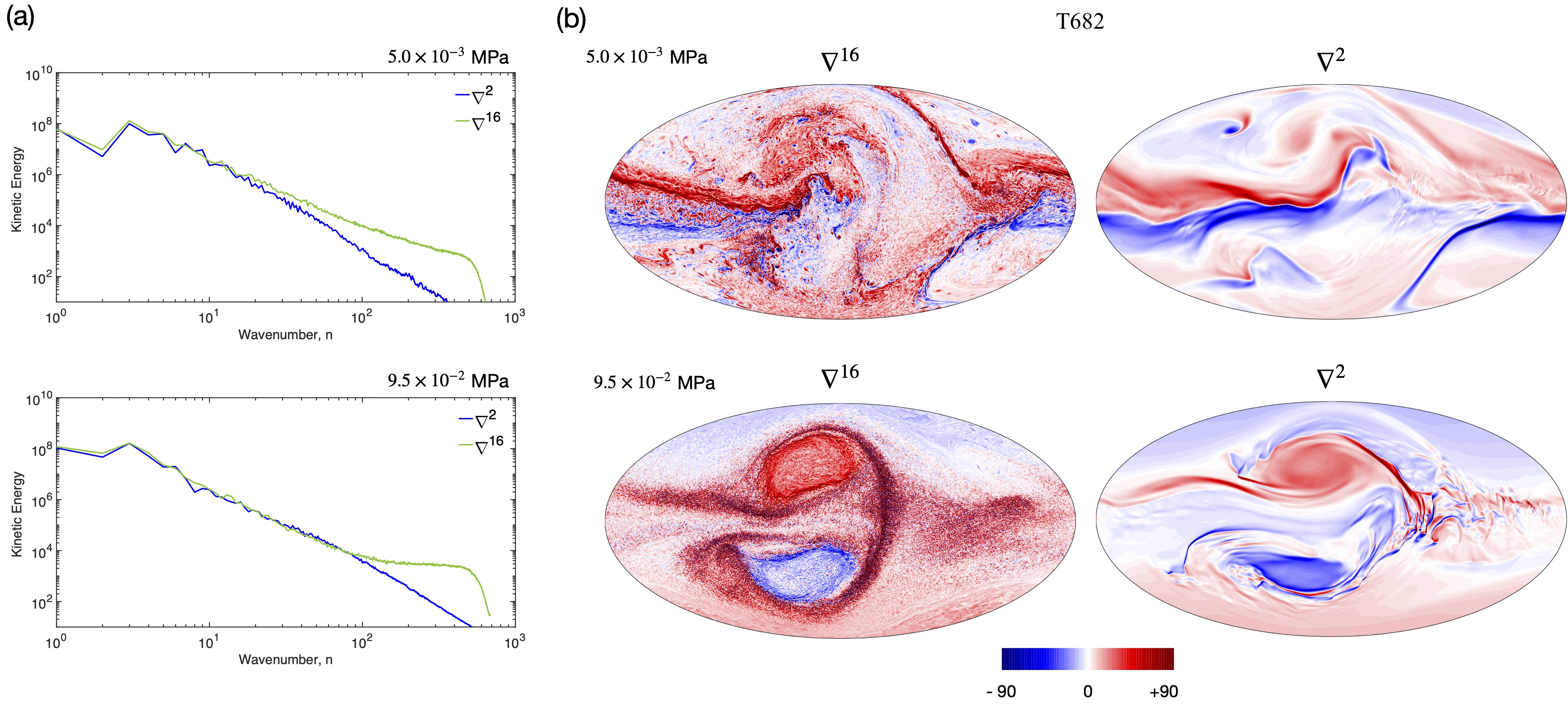}}
  \caption{Two T682L20 simulations, identical in all respects except for $\frakp
    \in \{1, 8\}$ (and the correspondingly adjusted $\nu_{2\frakp}$
    coefficients, $\nu_{2} = 1.0\! \times\! 10^{-8}$ and $\nu_{16} = 2.3\!
    \times\! 10 ^{-48}$).  The kinetic energy spectra (a) and the corresponding
    relative vorticity fields (b) at $t = 80$.  Both simulations contain similar
    energy contents up to $n \approx 80$ at $p = 9.5 \times 10^{-2}$ and only
    up to $n \approx 15$ at $p = 5.0 \times 10^{-3}$.  With $\frakp = 1$, energy
    is excessively dissipated from high wavenumbers and the relative vorticity
    fields are devoid of small scale dynamics.  While the energy contents of
    large-scales are similar in both simulations, $\frakp =1$ simulation
    contains noticeably less energy at $n = 2$ (at both $p$-levels).  At $p =
    5.0 \times 10^{-3}$, the $\frakp = 8$ simulation exhibits a dynamic modon,
    but the $\frakp = 1$ simulation does not capture the modon.  At $p =
    9.5\times 10^{-2}$, the modon is noticeably weaker and more sluggish in the
    $\frakp =1$ simulation.}
  \label{fig:T682_del2del16}
\end{figure*}

The unphysical consequence of the above over-dissipation ({\it particularly at
  the high $n$}) is more clearly seen with T682L20 simulations, presented in
Fig.~\ref{fig:T682_del2del16}.  In the figure, two simulations with different
$\frakp \in \{1, 8\}$ are presented (with accordingly adjusted $\nu_{2\frakp}$):
$(\frakp,\nu_{2\frakp}) \in \{\,(1,\, 1.0\!\times\! 10^{-8}),\ (8,\, 2.3\!
\times\!  10 ^{-48})\,\}$; otherwise, the two simulations are identical.  The
kinetic energy spectra and the corresponding relative vorticity fields at $t\!
=\!  80$ are shown.  At $p = 9.5\! \times\! 10^{-2}$, both simulations contain
similar amounts of energy up to only $n \approx 80$ and, even more unsettlingly,
only up to $n \approx 15$ at $p = 5.0 \!\times\! 10^{-3}$.  The latter is only
$\sim$2\% of the available range of $n$ -- i.e. nearly 98\% of the simulation is
subjected to over-dissipation.  Hence, in the $\frakp = 1$ simulation,
small-scale features are suppressed in the vorticity fields (at both $p$-levels
shown).  Note that, while the energy content in the large-scales (e.g. $n \lsim
15$) is similar in the two simulations, the $\frakp =1$ simulation noticeably
show less energy in the $n = 2$ mode (which correspond to modons) at both
$p$-levels.  Consequently, in the $\frakp = 1$ simulation, the modon is not
present at $p = 5.0\!\times\! 10^{-3}$ (cyclones are detached) and is
comparatively much weaker (than in the $\frakp = 8$ simulation) at $p =
9.5\!\times\! 10^{-2}$.

\subsection{Dissipation Order}\label{sec:order}

As might be expected from the previous sub-section, dissipation order~$\frakp$
strongly affects convergence.  And, there exists a `lower bound' on $\frakp$
for convergence: for the setup used in this paper, $\frakp > 4$ is required
(along with $\mbox{T} \ge 341$).  This is shown explicitly in
Fig.~\ref{fig:delcomp}, wherein simulations at T341 resolution are presented
with identical setup -- except for $\frakp$ (and, correspondingly,
$\nu_{2\frakp}$).  Recall that a rational procedure is used to obtain the value
of $\nu_{2\frakp}$, resulting in the same dissipation rate at $n = \mbox{T}$ for
all $\frakp \in \{1, 2, 4, 8\}$.  This gives $(\nu_2, \nu_4, \nu_8, \nu_{16}) =
(10^{-3}, 10^{-8}, 10^{-18}, 10^{-43})$: these values ensure equatable
comparisons between simulations with different $\frakp$
\citep[e.g.][]{ChoPol96a,ThraCho11,PoliCho12}.  As before, convergence is
diagnosed in three ways -- i.e. the instantaneous $\hat{\cal E}$ at $p =
0.095$~(Fig.~\ref{fig:delcomp}a), the $\zeta$-field from which the corresponding
spectrum in Fig.~\ref{fig:delcomp}a is obtained~(Fig.~\ref{fig:delcomp}b) and
the $\phi$--$t$ Hovm\"{o}ller plot of the $\zeta$-field at $p = 0.095$ from the
simulations presented in Figs.~\ref{fig:delcomp}a and
\ref{fig:delcomp}b~(Fig.~\ref{fig:delcomp}c).  Note that the different times of
the frames in Figs.~\ref{fig:delcomp}a and~\ref{fig:delcomp}b (shown at the
bottom of each frame in the latter) are chosen so that the modon, which forms in
all the simulations in the figure, is located closest to the planet's
sub-stellar point -- to adjust for the `phase shift' in the modon’s position in
physical space: recall that such a phase shift does not affect $\hat{\cal E}$.
The plots in Fig.~\ref{fig:delcomp}c show the differences in time, as well as in
space.

In Fig.~\ref{fig:delcomp}a, $\hat{\cal E}$ for $\frakp \in \{1, 2\}$ are
dissipated much more strongly compared with that for $\frakp \in \{3,4\}$ --
particularly at high $n$.  In fact, the over-dissipation is noticeable already
starting at $n \approx 40$, even with $\frakp\! =\! 2$.  Accordingly,
Fig.~\ref{fig:delcomp}b shows that the modons are significantly diffused and
small-scale structures are also noticeably absent, when the $\frakp \in \{1,
2\}$ fields are compared with the $\frakp \in \{4, 8\}$ fields.  Given this, we
expect current extrasolar planet simulation studies which use $\frakp \in \{1,
2\}$ dissipation order to display a much narrower `energetically-significant'
spectral range and, importantly, much less dynamic large-scale flow and
temperature structures.  Even with $\frak{p}\! =\! 4$, energy content is greatly
depleted for $n \ga 50$.  For $\frakp\! =\! 8$, notice the dissipation range
(super-exponentially decaying region at high~$n$) beginning at $n = n_{d(16)}
\approx 270$, signifying the presence of a proper conduit for energy and
potential enstrophy effluxes at the length-scale of T; here $n_{d(2\frakp)}$ is
the `(hyper)dissipation wavenumber'.  In contrast, a dissipation range is either
only weakly present or not at all present for the other $\frakp$ values.
Qualitatively, the above behaviours are generic and not restricted to a
particular $p$-level or resolution (provided that the latter is at least T341).

Fig.~\ref{fig:delcomp}b shows the non-convergence behaviour in physical space.
The behaviour is again consistent with that encountered in the corresponding
spectral space (Fig.~\ref{fig:delcomp}a).  In Fig.~\ref{fig:delcomp}b, the flow
field of the $\frakp = 1$ simulation is smooth almost everywhere.  Both the
cyclonic and anticyclonic modons that form are very diffused; and, the stronger,
cyclonic modon executes a simple, quasi-steady translation in the westward
direction.  In the flow field of the $\frakp = 2$ simulation, the cyclonic
modon's motion is more energetic and even slightly chaotic; its two constituent
cyclones can spread further apart in latitude than in the $\frakp = 1$ case, as
seen in the frame.  In the flow fields of the $\frakp \in \{4,8\}$ simulations,
the modon's motion and environment are much more complex than in the $\frakp =
2$ simulation; in particular, it generates a large quantity of small-scale
vortices at its periphery and in the equatorial region, primarily to its east
(see also Fig.~\ref{fig:refsol}b, right).  The principle differences between the
$\frakp \in \{4,8\}$ flow fields are more quantitative, rather than qualitative.
For example, with $\frakp = 8$ the modon generates many hundreds\footnote{The
  number is resolution dependent and can reach up to thousands at T682
  resolution (Fig.~\ref{fig:T682_del2del16}).} of small-scale vortices at its
periphery, compared to many tens with $\frakp = 4$.  Never the less, this leads
to a noticeable difference in the {\it evolutions} of $\frakp \in \{4,8\}$
simulations, as will be seen shortly: with $\frakp = 8$ the modon is more robust
and its motion is more chaotic.  The general picture of these simulations is
consistent with the noticeably higher energy content of the high wavenumbers in
higher~$\frakp$ simulations, seen in Fig.~\ref{fig:delcomp}a.  Unsurprisingly,
the general picture is also consistent with the behaviour in fully turbulent
simulations reported by \citet{ChoPol96a}, due to the turbulence produced by
{\it a}geostrophy here.

As just alluded to above, the non-convergence behaviour extends beyond a single
time frame: it persists over a long time, as shown in Fig.~\ref{fig:delcomp}c.
In fact, it persists over the duration of the simulation, after the initial
ramp-up period of $\sim$10~days.  Broadly, the $\frakp \in \{4,8\}$ simulations
can be grouped apart from the $\frakp \in \{1,2\}$ simulations according to
their evolutions: the simulations in the former group evolve qualitatively
similar to each other, while the simulations in the latter group evolve
qualitatively different than the simulations in the former group -- as well as
from each other (as already discussed above).  In addition to the $\frakp = 1$
flow field being essentially smooth everywhere over the entire duration, the
evolution is highly periodic; this is caused by the traversal of the cyclonic
modon around the planet with a period of $\approx$\,8~days.  In the $\frakp = 2$
simulation, the modon is more chaotic and short lived, with its constituent
cyclones ultimately detaching from each other.  In contrast, in the $\frakp \in
\{4,8\}$ simulations, the modon is much more robust and its life-cycle is much
more complex.  That is, the cyclonic modon initially forms $\sim\! 40^\circ$ to
the west of the sub-stellar point, as in the $\frakp \in \{1,2\}$ simulations;
but, then it proceeds to oscillate back and forth between the western terminator
and a point to the east of the sub-stellar point -- at times `hanging' on the
day-side for up to $\sim\!  17$~days.  Throughout this phase, the modon also
generates up to many hundreds small-scale vortices -- later disintegrating into
additional small-scale vortices and subsequently reforming (e.g. $t \approx
83$).  The entire cycle repeats often.  Over the duration presented in
Fig.~\ref{fig:delcomp}c, the $\frakp \in \{4,8\}$ evolutions both exhibit
$\sim$14~life-cycles of a modon forming on the day-side and traversing the
planet, with an average period of $\sim$12~days (cf. period of $\approx$\,8~days
in the $\frakp \in \{1,2\}$ evolutions).  The main qualitative difference
between the $\frakp \in \{4,8\}$ evolutions are the phase difference and the
cycle durations.

\begin{figure}
  \centerline{\includegraphics[scale=0.26]{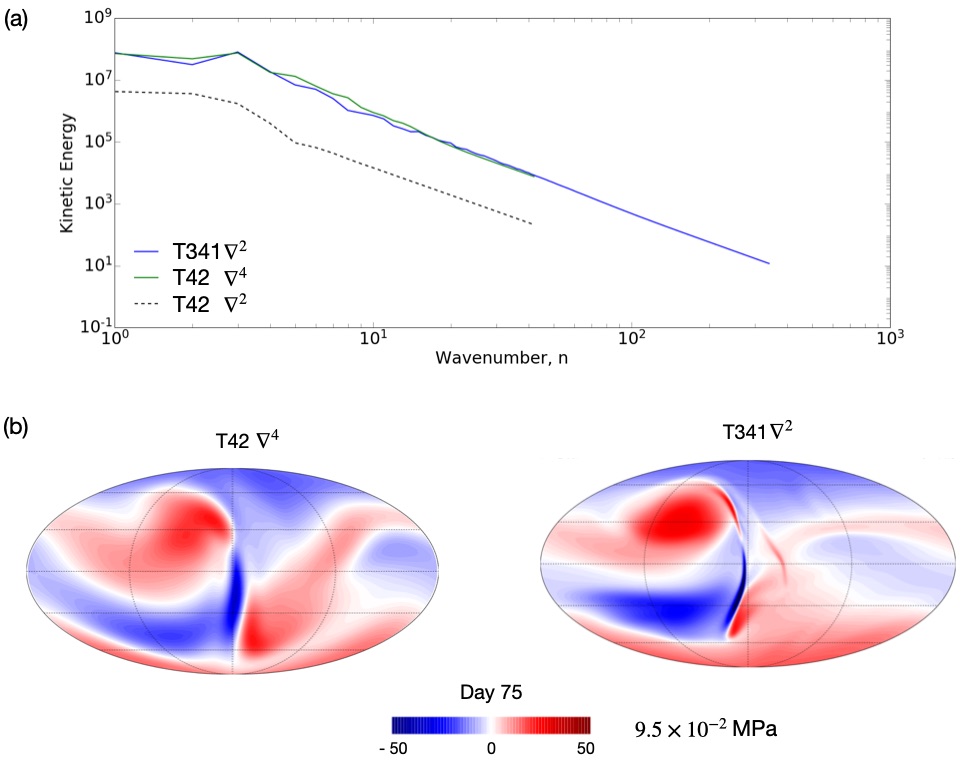}}
  \caption{Kinetic energy spectra (a) and flow field (b) from simulations with
    $\frakp \in \{1,2\}$ and two resolutions, T42L20 and T341L20, at $t = 75$
    and $p = 9.5\!\times\! 10^{-2}$.  The T42 simulation using the higher order
    dissipation operator ($\frakp = 2$) permits the spectrum to be much closer
    to that of a T341 resolution simulation with a lower order dissipation
    operator ($\frakp = 1$).  The flow fields in these two simulations are much
    more similar to each other than that from the T42 simulation using the
    $\frakp = 1$ operator (cf. Fig.~\ref{fig:del2rez}b) -- showing the utility
    of hyper-dissipation, even with lower resolution.}
 \label{fig:T42_T341}
\end{figure}

Thus far, we have motivated the use of high-order dissipation in attaining
convergence.  However, we wish to underscore the point that hyperdissipation is
more broadly an indispensable element in circumstances when the cost of
effecting simulations with adequate resolutions is nearly prohibitive, as in
extrasolar planet simulations.  Consider Fig.~\ref{fig:T42_T341}, for example.
It demonstrates how {\it a low resolution simulation employing a high-order
  dissipation can `emulate' the $\hat{\cal E}$ of a much higher resolution
  simulation employing a low-order viscosity} -- up to nearly $n_{d(2\frakp)}$
of the lower resolution simulation; recall that $n_{d(2\frakp)}$ is the fiducial
dissipation scale $n_d$ for the given $2\frakp$.  This ability is a tremendous
practical advantage.  In the figure, two T42 simulations with $\frakp = \{1,2\}$
and $(\nu_2,\nu_4) = (10^{-7},10^{-9})$ are compared with a T341 simulation with
$\frakp = 1$ and $\nu_{2} = 10^{-8}$.  In the latter simulation, $\Delta t$ is
reduced by 8 times to accommodate the higher T value, but all three simulations
are identical otherwise.  The time of all the plots in the figure is $t = 75$.

In Fig.~\ref{fig:T42_T341}a, notice how increasing the dissipation order of the
lower resolution simulation permits it to achieve a similar level of
inviscidness -- up to $n = \mbox{T}$ in this illustration (cf. green and blue
full lines).  In the figure, the T42 simulation with $\frakp = 1$ from
Fig.~\ref{fig:del2rez} is reproduced for reference (black dotted line).
Comparing this simulation to the T341 simulation with the same $\frakp$ and
$\nu_2$ (blue line), the T42 simulation is much more dissipative with the
large-scale flow completely devoid of dynamism: if convergence were achieved,
$\hat{\cal E}$ from the two simulations would be nearly the same up to $n
\approx n_{d(2)}$ (as in the blue and green lines).  We emphasize that the T42
simulation with $\frakp =1$ here is comparable to most current extrasolar planet
atmosphere flow simulations in terms of effective resolution.  Unfortunately,
employing high $\frakp$ operator at T42 does {\it not} lead to convergence in
extrasolar planet simulations -- as discussed in section~\ref{sec:rez}.
Fig.~\ref{fig:T42_T341}b shows the flow fields from which two of the spectra in
Fig.~\ref{fig:T42_T341}a (T42 with $\frakp = 2$ and T341 with $\frakp = 1$) have
been obtained.  The two flow fields in Fig.~\ref{fig:T42_T341}b are similar (but
not identical), here again consistent with the spectra in
Fig.~\ref{fig:T42_T341}a.  For example, the cyclonic modon that forms west of
the sub-stellar point begins to decouple when it reaches near the western
terminator in both simulations.  The T42 field here should be compared with the
T42 field of Fig.~\ref{fig:del2rez}, in which the modon is much weaker (N.B. the
difference in scale of values) and the flow is north--south symmetric.

\subsection{Deep Atmosphere}\label{sec:deep}

\begin{figure*}
  \centerline{\includegraphics[scale=0.12]{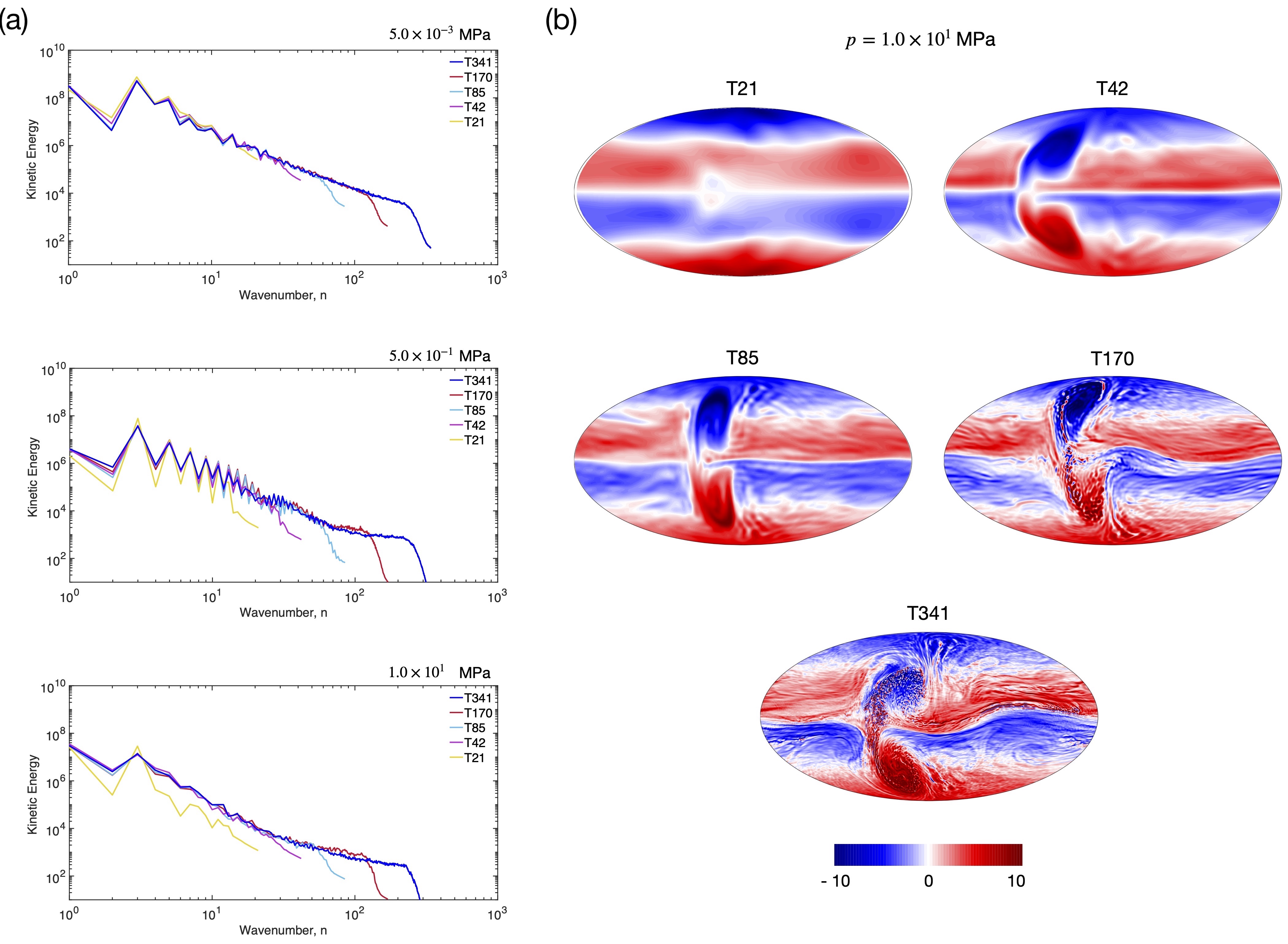}}
  \caption{`Deep atmosphere' simulations with vertical domain range $p \in
    [0,10]$ and L200 vertical resolution at different horizontal resolutions.
    Simulations are identical in all respects (except horizontal resolution and
    their correspondingly adjusted $\nu_{16}$ and $\Delta t$ values); $\frakp =
    8$ in all the simulations shown.  Kinetic energy spectra (a) at the
    $p$-levels: $5.0\times 10^{-3}$, $5.0\times 10^{-1}$ and $1.0 \times 10^{1}$
    at $t\! =\!  45$ and the instantaneous $\zeta$-fields from which the spectra
    in (a) were obtained.  At T341 resolution, simulations are not spectrally
    converged and the non-convergence behaviour increases towards the bottom of
    the domain.  Both traits are expected and broadly similar to the shallow
    atmosphere case (cf. Figs.~\ref{fig:specrez}b and \ref{fig:del2rez}a).  The
    corresponding flow fields in (b) at $p = 1.0 \times 10^{1}$ are markedly
    different above T170L100.  The flows below this resolution are much less
    dynamic.}
 \label{fig:deep_spec}
\end{figure*}

\begin{figure*}
  \centerline{\includegraphics[scale=.6]{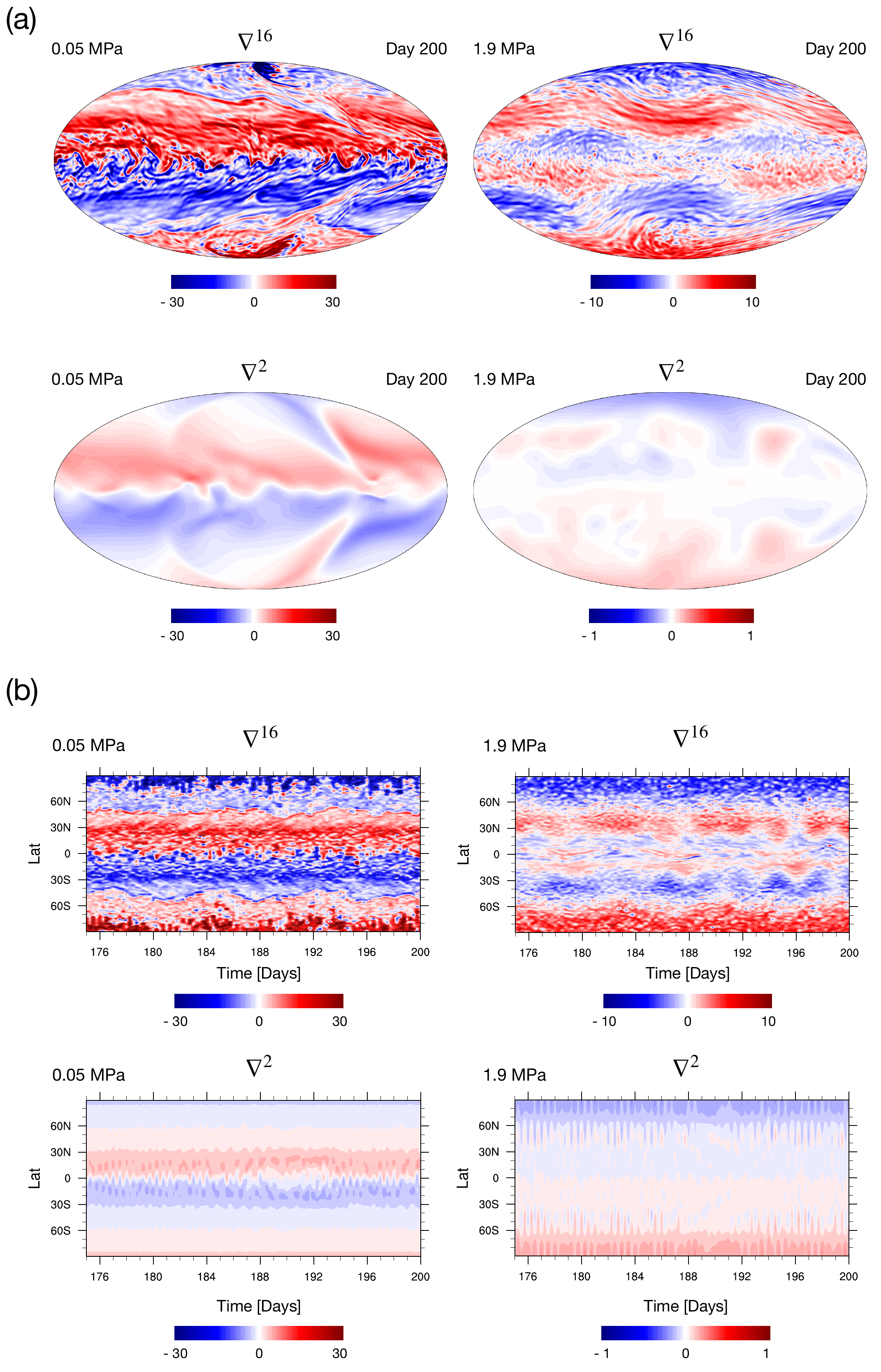}}
  \caption{Relative vorticity fields in Mollweide projection at $t = 200$ for $p
    \in \{0.05,1.90\}$ from the T170L200 deep atmosphere simulations.  The
    simulations are set up identically -- except for their dissipation order,
    $\frakp \in \{1, 8\}$ (and their correspondingly adjusted viscosity
    coefficient $\nu_{2\frakp}$).  Red (blue) colour represents positive
    (negative) values of relative vorticity.  Deep atmosphere flow fields are
    more zonal compared to shallow atmosphere flow fields at the $p$-levels
    common to both atmospheres (i.e. $p \in [0.0,0.1]$), particularly in the
    equatorial region and at the levels away from the $p_{\rm top}$.  Also, the
    large-scale flow is now separately barotropic in two distinct vertical
    sub-regions, $0.05 \ga p \ga 1$ and $1 \ga p \ga 10$; in the latter region,
    the flow is much {\it less} zonal than in the former region.  This general
    behaviour is independent of~$\frakp$.  However, the amplitude of the
    equatorial jet (the roughly 60$^\circ$--wide band of $\zeta$ with the
    transition in sign across the equator) is greater for higher $\frakp$; here
    the jet amplitude is related to the gradient of $\zeta$ contours.}
  \label{fig:deep_field}
\end{figure*}

Thus far, we have laid focus on the convergence behaviour of a shallow
atmosphere, for which $p_{\rm bot}$ is set to be 0.1.  As mentioned in
section~\ref{sec:method}, the planetary radius $R_p$ is typically measured out
to this $p$-level on giant planets: importantly, it is also {\it roughly the
  $p$-level at which the optical path length of visible and near infrared
  radiation, entering from the top, begins to reach unity}
\citep[e.g.][]{Irwi09}.  For Jupiter, a $p_{\rm bot}$ range of 0.1 to 1.0 is
common for studying the dynamics near the visible cloud deck level
\citep[e.g.][]{VasaShow05,Sancetal19}.  This is despite the expectation that the
radiatively stratified region extends to a greater $p$-level (i.e. deeper in),
with the actual value currently uncertain.  For extrasolar planets, the choice
of $p_{\rm bot}$ value is more arbitrary and arguable.  However, we have
observed that a number of atmospheric flow and general circulation properties of
Jupiter-like extrasolar planets -- including convergence -- is independent of
the value of $p_{\rm bot}$ (up to $p_{\rm bot}= 20$).  In this paper, we
continue to emphasize the generic properties of convergence that apply to both
shallow and deep atmospheres.

Because the existence -- and, if so, the location of -- a solid surface for
giant planets is unknown, the vertical domain range of simulations is generally
chosen based on the physical phenomenon of interest and the computational
resources available.  A natural, and physically instructive, choice in this
situation is to set $p_{\rm bot}$ to the level where there is a sharp change of
vertical gradient or a jump in the basic stratification (e.g. laterally- and
temporally-averaged Brunt-V\"{a}is\"{a}l\"{a} frequency or density).
Unfortunately, information on the detailed basic stratification structure is
also unavailable.  Formally, because of the restriction to the large scales
imposed by the hydrostatic balance condition, equations~(\ref{eq:pe}) are
strictly valid only for the stably-stratified radiative region, which overlies
the unstably-stratified convective region.  On large parts of the day side, the
boundary between these two stability regions could be located at a depth greater
than $p = 0.1$ because of the intense irradiation from the planet's host star.
The precise depth depends on $\lambda$ and $\phi$.  Simple, one-dimensional
models predict a value for the sub-stellar point as large as $p \sim\!  1000$
\citep[e.g.][]{GuilShow02}.

In the present sub-section, we consider the deep atmosphere, in which $p_{\rm
  bot}$ is 1.0 or 10.  Our discussion here centres mainly on the latter
instantiation because, as pertains to convergence, there is no qualitative
difference between simulations with $p_{\rm bot} \in \{0.1, 1.0\}$.
Quantitatively, the general flow pattern changes monotonically as $p_{\rm bot}$
increases from 0.1 to 1.0: modons become weaker and wider and the equatorial jet
becomes more zonal, as $p_{\rm bot} \rightarrow 1$.  The qualitatively-robust
general behaviour is due to the strongly barotropic quality of the flow, which
is intimately related to the setup used.  \citet{PoliCho12} have previously
reported quantitative changes in the baroclinicity when the $p_{\rm bot}$ is
similarly increased in their idealised study of jet instability on hot-Jupiters.
In that study, the (sectoral) wavenumber of the gravest unstable mode decreases
slightly with a corresponding slight increase in the resulting flow's zonality.
Here the value of $p_{\rm top}$ in all the deep simulations is same as in the
shallow atmosphere simulations, to facilitate unambiguous comparisons.  A
detailed discussion of the effects of vertical range variation will be provided
elsewhere.  We note that, in the employed setup, the depth at which the
atmosphere ceases to be thermally forced is $p = 1.0$; in addition, the forcing
is very weak in the region $0.1 \ga p > 1$, compared to the levels near $p_{\rm
  top}$.

In light of the very high horizontal resolution requirement for convergence in
$p_{\rm bot} = 0.1$ simulations and the behaviour expected in $p_{\rm bot} = 10$
simulations based on the similarity of $p_{\rm bot} \in \{0.1,1.0\}$
simulations, adequate vertical resolution for the $p_{\rm bot} = 10$ deep
atmosphere is at present computationally nearly prohibitive: we expect that
effectively a series of long-duration, T341L2000 (or higher) simulations is
needed for a robust assessment of full convergence for the atmosphere with
$p_{\rm bot} = 10$ (and even greater~L for larger $p_{\rm bot}$).  None the
less, a `practical assessment' can still be carried out with a reduced vertical
range and/or layer density~$\ell$ (i.e.~L {\it per}~MPa).  As with the $p_{\rm
  bot} = 1$ simulations, the $p_{\rm bot} = 10$ simulations discussed in this
paper (which are with up to T341L200 resolution and $t \in [0,500]$ duration)
exhibit behaviours that are quantitatively different than the $p_{\rm bot} \in
\{0.1, 1.0\}$ simulations -- particularly, when $\ell$ is high (e.g. $\ell \ga
200$) in the $p_{\rm bot} \in \{0.1, 1.0\}$ simulations.  This is in part due to
the reduction of vertical resolution in the domain's upper region (where the
forcing is the strongest) in the $p_{\rm bot} = 10$ simulations.  However, we
also observe weakening of the modons and strengthening of the overall flow's
zonality when L is increased while $\ell$ is held fixed (i.e. when layers are
simply added at the bottom, keeping the vertical resolution in the domain's
upper region fixed).  The latter behaviour is principally due to the
aforementioned barotropic nature of the flow, in which the overall flow
structure is in effect stretched vertically down to $p_{\rm bot}$, and the
specified thermal forcing that must now drive a much larger mass of atmosphere
than in the shallower atmospheres.

As mentioned, the horizontally (as well as vertically) under-resolved $p_{\rm
  bot} = 10$ simulations are still revealing for convergence purposes.  For
example, consider Fig.~\ref{fig:deep_spec}.  It shows the $\hat{\cal E}$ at $t =
45$ from a series of simulations -- all with $(p_{\rm bot}, \mbox{L}, \frakp) =
(10, 200, 8)$, but with different T values.  The spectra from the flow fields at
the levels, $p = (5.0 \times 10^{-3}, 5.0 \times 10^{-1}, 1.0 \times 10^1)$, are
shown in the (top, middle, bottom) panels in Fig.~\ref{fig:deep_spec}a.  First,
note the nearly identical spectra (up to the dissipation scale $n_{d}$ of each
resolution given) for all T at $p = 5.0 \times 10^{-3}$; this match among the
spectra here is actually slightly better than in the shallow atmosphere
simulations (cf. Fig.~\ref{fig:specrez}).  However, the spectra at the $p =
\{5.0 \times 10^{-1}, 1.0 \times 10^1\}$ levels again exhibit spectral blocking
(see e.g. the T170 spectra).  This is very reminiscent of the behaviour already
encountered in the shallow atmosphere simulations (cf.  Fig.~\ref{fig:specrez}).
Note also the much stronger oscillations in the lower $n$ part of the spectra at
$p = 5.0 \times 10^{-1}$, which suggest a much stronger zonality at that
$p$-level; such pronounced, `mid'-level feature is not present in the $p_{\rm
  bot} \in \{0.1, 1.0\}$ simulations.  Most importantly, as expected from what
we have already observed in the shallow atmosphere case, the deep atmosphere
simulations are not converged even at T341L200 resolution -- particularly away
from the top levels of the domain.  This can be seen in
Fig.~\ref{fig:deep_spec}b, in which the flow complexity clearly increases with T
(and dynamism only with $\mbox{T} \ge 170$).  In sum, we reiterate the following
salient point: {\it regardless of the vertical range of the modelled atmosphere,
  at least T341 horizontal resolution is needed for convergence}.

Unsurprisingly, the above general behaviour with varying T is reproduced in the
deep simulations with varying $\frakp$.  This is shown in
Fig.~\ref{fig:deep_field}, in which the instantaneous $\zeta$-fields
(Fig.~\ref{fig:deep_field}a) and their corresponding Hovm\"{o}ller plots
(Fig.~\ref{fig:deep_field}b) are presented; here all the simulations in the
figure are at T170L200 resolution.  In the figure, several features are readily
noticeable.  First, as already discussed, deep atmosphere flow fields are more
zonal compared to shallow atmosphere flow fields at the $p$-levels common to
both atmospheres -- particularly in the equatorial region, and at the levels
away from the $p_{\rm top}$ (see Fig.~\ref{fig:deep_field}a).  Second, in
$p_{\rm bot} = 10$ simulations, the flow is also strongly barotropic, but
roughly in two vertical sub-regions: $0.05 \la p \la 1$ and $1 \la p \la 10$.
This is expected, given the specified forcing structure\footnote{Recall that the
  $p \ge 1$ region is not thermally forced in the setup.}; it is also broadly
consistent with behaviours reported in \citet{ThraCho10} and \citet{PoliCho12}.
Third, in the latter region, the flow is generally much less zonal than in the
former region.  The azonal behaviour is supported by the lower boundary and is
observed in all simulations, regardless of the $p_{\rm bot}$ value.  The above
features are independent of~$\frakp$.  However, the field amplitude for $\frakp
= 1$ is globally low compared to those from the simulations with higher
$\frakp$.  Hence, the amplitude of the equatorial jet (the roughly
60$^\circ$--wide band of $\zeta$ with the transition in sign across the equator)
is greater for higher $\frakp$.\footnote{Note that the jet amplitude is related
  to the gradient of $\zeta$ contours.}

The $\frakp = 8$ flow field is also distinct from the $\frakp \in \{2,4\}$ flow
fields (not shown), similar to what was observed in Fig.~\ref{fig:delcomp} for
the shallow atmosphere.  As in the shallow atmosphere simulations, the higher
$\frakp$ field contains many more small-scale flow structures.  But, in the deep
atmosphere simulations, the small-scale vortices are associated more strongly
with a process of continuously peeling-off from the dynamic equatorial jet --
rather than from the combined action of modon radiation and jet instability,
seen in the shallow atmosphere simulations.  Also clearly visible in the flow
fields (but, importantly, not in the spectra) is the large-amplitude Rossby wave
at the jet's core (undulation of the $\zeta \rightarrow 0$ line in both of the
$\frakp =\{1,8\}$ simulations, at both of the $p$-levels shown).  The
undulations aids in the production of large-scale, as well as small-scale,
vortices.

The flow of the deep atmosphere simulations also evolves very differently than
that of the shallow atmosphere simulations.  This can be readily seen in the
Hovm\"oller plots shown in Fig.~\ref{fig:deep_field}b.  The duration in the
plots is $t \in [175,200]$, well after equilibration of the bottom region (which
occurs much later than in the shallow atmosphere simulations).  In
Fig.~\ref{fig:deep_field}b, the difference in behaviours of $\frakp \in \{1,8\}$
simulations seen in Fig.~\ref{fig:deep_field}a persists in time.  Here
large-amplitude Rossby wave {\it propagation} can be seen more clearly
(indicated by undulations of $\zeta$-amplitudes near the equator in time).  In
addition, the $\frakp = 8$ evolution appears to contain one or two more jets
than in the $\frakp = 1$ evolution.  In the latter, large time-scale variation
of period $\sim$\,5~days can be seen at $p = 0.05$, as well as shorter
time-scale variations of period $\sim\! 1/3$ day at both of the $p$-levels
shown.  For the modelled planet atmosphere, because ${\cal L_R} / R_p = {\cal
  O}(1)$, where ${\cal L_R} \equiv (g {\cal H} / \Omega^2)^{1/2}$ is the
external Rossby deformation scale \citep[e.g.][]{Holt04}, the number of bands
(jets) is expected to decrease over long time -- especially at the greater
$p$-level.  This is because ${\cal L_R}$ is effectively the interaction length
between the jets \citep[see][]{ChoPol96a,ChoPol96b,Choetal08}.

We have also observed several additional features worthy of mention.  First,
little difference in the basic structure is discernible between the $\frakp \in
\{2,4\}$ evolutions, while the $\frakp = 1$ evolution is markedly different
(most noticeably in the amplitude) compared to the other two evolutions; on the
other hand, while the $\frakp = 8$ evolution is broadly similar to the $\frakp
\in \{2,4\}$ evolutions, vortices are smaller and the cross-equatorial flow is
much stronger in the former.  The latitudinal gradients are much sharper in the
$\frakp = 8$ evolution as well; hence, we expect correspondingly sharper jets.
After the initial ramp-up period, these features persist over the entire
duration of the simulations (up to $t = 500$).  As discussed, T341L200
simulations with $\frakp = 8$ and vertical domain range $p \in [0,1]$ show a
much more dynamic evolution, which is closer to the behaviour seen in
Fig.~\ref{fig:delcomp} for a T341L20 simulation with range $p \in [0,0.1]$.
Additionally, while the general jet behaviour is robust, the differences in the
jet core (with span $\phi \la |\!\pm 30^\circ|$) and the jet flanks (located at
$\phi \sim |\!\pm 30^\circ|$) is less pronounced in the T85L200 and T42L200
simulations (not shown), suggesting a transition similar to what was observed in
the shallow atmosphere simulations at the T341 resolution
(cf. section~\ref{sec:rez}).

In summary, the non-convergence of simulations with up to T341L200 and $\frakp =
8$ as well as the behavioural trends laid out in the foregoing discussion
together indicate convergence could be achieved at the next higher horizontal
resolution\footnote{The resolution is generally chosen from the truncation
  wavenumber set, $\mbox{T} \in \{21, 42, 63, 85, 106, 170, 341, 682,
  1364,\,\hdots\}$, which permits the most efficient use of the fast Fourier
  transform algorithm employed in the code; specifically, the length of the
  transforms must be a number greater than 1 that has no prime factors other
  than $\{2,3,5\}$ \citep{Temp92}.} -- {\it if} the layer density $\ell$ is much
greater (e.g. $\ell \sim 200$, in contrast to 20 above).  Given this, we
estimate at least T341L2000 resolution with $\frakp = 8$ is needed for a robust
assessment of convergence (and preferably a resolution of T341L4000).  Even with
such a resolution, the simulations are not likely to be converged away from the
top region of the modelled domain -- as we have shown here in both the shallow
and deep atmosphere context.

\section{Conclusion}\label{Conclusion}

In this paper, we have summarised the main results from an in-depth exploration
of convergence in extrasolar planet atmosphere flow simulations.  Converged flow
dynamics solutions are critical because they form the backbone for accurate
modelling of other important (e.g. thermal, radiative and chemical) processes
which directly affect observations and interpretations.  In short, we have found
that a horizontal resolution of T341 with a $\grad^{16}$ hyperdissipation
(roughly equivalent to at least a $2000 \times 1000$ finite-difference grid,
when solutions are smooth\footnote{A solution $\xi$ is sufficiently smooth if it
  satisfies the Lipschitz condition: $|\xi(\bfx,t) - \xi(\bfx^\prime,t)| \le
  c\,|\bfx - \bfx^\prime|$ for all $\bfx$ and $\bfx^\prime$ and $t \ge 0$, where
  $c > 0$ is a real constant \citep[e.g.][]{Krey78}; this condition is
  satisfied, for example, when $|| \nabla \xi || < \infty$, where $||\cdot||$ is
  the norm operator. }) and a concomitant vertical resolution of $\ell\! \sim\!
200$ (i.e., $\sim$200 levels, or layers, {\it per} MPa), is minimally needed for
convergence in hot-planet simulations: we suggest T341L4000.  Crucially, this is
because of the energetic small-scale vortices and waves, which naturally arise
in the {\it a}geostrophic condition of the modelled planet atmosphere.

More broadly, this work presents several significant implications for extrasolar
planet atmosphere studies.  First, given that we have only invoked the
conditions of stratification/density jump and {\it a}geostrophy, the results
here also apply to simulations of close-in telluric planet atmospheres
(ostensibly away from the boundary layer) -- if similar forcing and initial
condition are used.  In fact, the results are arguably more appropriate for
telluric planets in some ways because of the inescapable lower boundary required
by discretization in simulation work.  Of course, for such planets the boundary
conditions should be augmented (e.g. with `no-slip' and/or prescribed
temperature at the bottom surface).  Second, the results here also apply to
studies that solve the full Navier--Stokes equations.  This is because {\it
  a}geostrophy poses the same numerical difficulty (e.g. resolving small-scale
flow structures) for both the primitive and Navier--stokes equations.  Finally,
the above resolution requirement suggests that current extrasolar planet
atmosphere flow simulations are not converged.  To the best of our knowledge,
simulations employing the same (or similar) setup have thus far been performed
with a lower resolution {\it and} dissipation order.  Indeed, our results
suggest that current simulations are erroneously converging to unphysical
states: we recommend restricting the domain range and/or reformulating the
equations for a more appropriate vertical coordinate.

The atmosphere has a number of properties that make the numerical solution of
its governing equations especially challenging.  Most obvious is the sphericity
and anisotropy, the latter in both vertical (radial) and horizontal directions.
That is, gravity and Coriolis acceleration impose a strong restraint in the
vertical and meridional directions, respectively.  The resulting stratifications
induce the vertical length scales to be typically much smaller than the
horizontal length scales and, to a much lesser extent, meridional scales to be
smaller than the zonal scales.  Hence, the ratio of horizontal to vertical model
resolution and the lateral extent of the model domain should be carefully chosen
to capture these anisotropies correctly.  Because of the particularly strong
vertical--horizontal anisotropy, atmospheric models are almost invariably
constructed with a certain number of levels or layers in the vertical, with
essentially the same horizontal grid or expansion basis structure at each level.
Thus, we have discussed convergence naturally separated into horizontal and
vertical issues.

Another fundamental challenge in modelling the atmosphere is its strong
multi-scale property, in both space and time.  In terms of spatial energy
spectra, the largest scales are energetically dominant, but the spectra are
shallow -- implying that, whatever the resolution of an atmospheric model, there
exists a significant dynamical variability near the resolution limit (which
requires careful handling by numerical methods).  Moreover, there is still the
issue of unresolved scales, as the numerical models are still far from achieving
realistic Reynolds number; this must be represented by a sub-grid model (here by
hyper-viscosity).  In addition, the atmosphere also supports dynamics with a
huge range of timescales. Care is needed in modelling fast processes to ensure
that the numerical solution is stable (in fact, convergence formally refers to a
scheme/code which is both numerically consistent {\it and} stable).  In contrast
to geostrophic conditions, fast (gravity and sound) waves can be energetically
much more significant in {\it a}geostrophic conditions.  The chosen numerical
method and code must be able to represent these features accurately.  Also,
certain properties of the atmosphere evolve slowly, either in a Lagrangian sense
(e.g.  the moisture content of an air parcel in the absence of condensation and
evaporation) or in a global integral sense (e.g. the total angular momentum or
energy of the atmosphere).  These conservation properties must also be captured
accurately.  Note that some spectral blocking is almost inevitable in a long
time integration of a nonlinear system (unless the dissipation is large).

A spectral method will conserve energy to a very high accuracy and will
faithfully resolve a front or a shock with only two or three grid points across
the structure.  However, if these structures are poorly-resolved (or worse,
under-resolved), its accuracy is no better than other methods.
Energy-conserving and front-tracking schemes generate smooth solutions, but
these may be far from the true solution.  A turbulence calculation, such as one
for extrasolar planet atmosphere, is almost by definition poorly resolved.  In
the present case, even dealiasing does not cure an under-resolved flow (when it
is not suppressed by over-dissipation).  For example, when a front forms, the
solution is smooth for a finite time interval and then develops a jump
discontinuity which the code is not able to resolve (leading to, inter alia,
spectral blocking).  Frontogenesis happens very rapidly once the modons grow to
a reasonable size and strength -- which occurs almost on a 1-day time-scale in
the simulations.  Hence, poorly-resolved simulations are predestined to fail
convergence from the beginning of the simulation.

Finally, when the model top corresponds to the top of the atmosphere, it is
reasonable to assume that there is no vertical mass flux across the upper
boundary.  This condition can be artificial, however, even when the upper
boundary is formally placed at the top of the atmosphere, because of practical
limitations on vertical resolution.  At present, a simple but fully justifiable
way of handling the upper boundary is not available.  For a horizontally
discrete model, the effective height of the surface at the model’s lower
boundary may be higher than the actual height averaged over the grid box, and
presents an additional challenge.

In this work, we have shown that a more numerically-suitable setup is needed (if
not a more realistic one).  Nominally, this means a more `balanced' initial
condition, which would mitigate the front and small-scale wave generation.
Notably, what is not captured in lower resolution and lower order viscosity
simulations (with the current setup) is the dominance of dynamic modons on
close-in extrasolar planets and the modon's influence in redistributing
important fields (such as temperature and chemically- and radiatively-active
species), as well observable variability of the planets. Here we have mainly
focused on numerical convergence issues, as they relate to the flow dynamics.
Hence, we have not discussed the effects of resolution and dissipation order on
the temperature field.  Indeed we have found that the flow and the temperature
fields are intimately linked.  This link is discussed in detail elsewhere, in
two companion studies to the present work (Cho, Skinner \& Thrastarson,
submitted; Skinner \& Cho, prep).



\section*{Acknowledgements}

The authors thank Craig Agnor, Heidar Th. Thrastarson and Ursula Wellen for
helpful discussions, as well as the referee for helpful comments.  We
are grateful for the hospitality of James Stone and the Department of
Astrophysical Sciences, Princeton University, where some of this work was
completed.  J.W.S. is supported by the UK's Science and Technology Facilities
Council research studentship. {\it In memoriam} Adam Showman.

\section*{Data Availability}
The data underlying this article will be shared on reasonable request to the 
corresponding author.


\end{document}